\documentclass[a4paper,11pt]{article}

\usepackage[english]{babel}
\usepackage{amsmath}
\usepackage{amssymb}
\usepackage{multirow}
\usepackage{graphicx}
\usepackage{subfigure}
\usepackage{wrapfig}
\usepackage{geometry}
\usepackage{cite}
\usepackage{float}

\makeatletter
\def\captionof#1#2{{\def\@captype{#1}#2}}
\makeatother

\def\kmpi2{\stackrel{\circ}{M_{\pi}^2}}
\def\kmk2{\stackrel{\circ}{M_{K}^2}}
\def\kmeta2{\stackrel{\circ}{M_{\eta}^2}}

\def\kmp2{\stackrel{\circ}{M_{P}^2}}

\def\klmpi2{\stackrel{\circ}{\stackrel{\sim}{M_{\pi}^2}}}
\def\klmk2{\stackrel{\circ}{\stackrel{\sim}{M_{K}^2}}}
\def\klmeta2{\stackrel{\circ}{\stackrel{\sim}{M_{\eta}^2}}}

\def\mmu2{\mu^2}

\def\mpi2{M_{\pi}^{2}}
\def\mk2{M_{K}^{2}}
\def\meta2{M_{\eta}^{2}}

\def\fpi2{F_{\pi}^{2}}
\def\fk2{F_{K}^{2}}
\def\feta2{F_{\eta}^{2}}

\def\mo{\stackrel{\circ}{M}}

\begin{document}
\thispagestyle{empty}

\begin{flushright}
LPT-ORSAY/12-20\\
\end{flushright}

\vspace{\baselineskip}

\begin{center}
\vspace{3\baselineskip}
\textbf{\Large 
Topological susceptibility on the lattice\\
and the three-flavour quark condensate
}\\

\vspace{3\baselineskip}
{\sc V\'eronique Bernard$^{(a)}$, S\'ebastien Descotes-Genon$^{(b)}$ and Guillaume Toucas$^{(b)}$}\\

\vspace{0.7cm}
{\sl $(a)$ Institut de Physique 
Nucl\'eaire, CNRS/Univ. Paris-Sud 11 (UMR 8608),\\
91406 Orsay Cedex, France\\

$(b)$ Laboratoire de Physique 
Th\'eorique, CNRS/Univ. Paris-Sud 11 (UMR 8627),\\
{\sl 91405 Orsay Cedex, France}}
\vspace{3\baselineskip}

\vspace*{0.5cm}
\textbf{Abstract}\\
\vspace{1\baselineskip}
\parbox{0.9\textwidth}{
We reanalyse the topological susceptibility assuming the possibility of a significant
paramagnetic suppression of the three-flavour quark condensate and a correlated enhancement
of vacuum fluctuations of $s\bar{s}$ pairs. Using the framework of resummed $\chi$PT,
we point out that simulations performed near the physical point, with a significant mass hierarchy between $u,d$ and $s$ dynamical quarks, are not able to disentangle the contributions from the quark condensate and sea $s\bar{s}$-pair fluctuations, and that simulations with three light quark masses of the same order are better suited for this purpose. 
We perform a combined fit of recent RBC/UKQCD data on pseudoscalar masses and 
decay constants as well as the topological susceptibility, and we reconsider the determination of lattice spacings in our framework, working out the consequences on the parameters of the chiral Lagrangian. 
We obtain $(\Sigma(3;2{\rm GeV}))^{1/3}=243 \pm 12$ MeV 
for the three-flavour quark condensate in the chiral limit. 
We notice a significant suppression compared to the two-flavour 
quark condensate $\Sigma(2;2{\rm GeV})/\Sigma(3;2{\rm GeV})=1.51\pm 0.11$ and we confirm previous findings of a competition between leading  order and next-to-leading  order contributions in three-flavour chiral series.}
\end{center}

\clearpage


\newpage
\setcounter{page}{1}

\section{Introduction}

Over the recent years, a joint experimental and theoretical effort has provided new insight concerning low-energy dynamics of QCD, through the investigation of $\pi\pi$ (re)scattering in $K_{\ell 4}$~\cite{Colangelo:2001sp,DescotesGenon:2001tn,Batley:2007zz,:2009nv,Colangelo:2008sm,DescotesGenon:2012gv,wip2} and $K\to 3\pi$ decays~\cite{Batley:2005ax,Cabibbo:2004gq,Cabibbo:2005ez,Colangelo:2006va,Gasser:2011ju}. It turns out that,
in the $N_f=2$ chiral limit ($m_u=m_d=0$)~\cite{chpt-su2}, the breakdown of chiral symmetry (responsible for the presence of pseudoscalar Goldstone bosons corresponding to pions) is triggered by 
the quark condensate $\Sigma(2)=-\lim_{m_u,m_d\to 0} \langle \bar{u}u\rangle$.
The situation is not so clear for the pattern of chiral symmetry breaking in the 
$N_f=3$ chiral limit where $m_s$ also vanishes and kaons and $\eta$ become also Goldstone bosons~\cite{chpt-su3,Colangelo:2010et}. 
In particular, one would expect chiral order parameters (such as the quark condensate) to decrease as the number of massless flavours in the theory increases.
This suppression is expected to be more and more pronounced as more and more light quark masses are sent to zero due to a paramagnetic suppression from the vacuum fluctuations of sea-quark pairs~\cite{DescotesGenon:1999uh}. This effect is hinted at by the significant violation of the Zweig rule in the scalar sector, and should affect  three-flavour dynamics involving kaons and eta mostly, but not (or only mildly) the two-flavour dynamics of pions alone~\cite{DescotesGenon:2000di,DescotesGenon:2000ct}. 
Such a suppression would lead to a decrease of the condensate in the $N_f=3$ case compared to its value in the $N_f=2$ limit. An unambiguous test of the pattern of $N_f=3$ chiral symmetry breaking comes from lattice simulations, which have made significant progress in the recent years, due to many improvements  both at the conceptual and technical levels. A first avenue consists in computing the spectrum and/or the dynamics of light pseudoscalar mesons in order to extract the low-energy constants encoding the pattern of chiral symmetry breaking in the two chiral limits (for a recent overview, see ref.~\cite{Colangelo:2010et}).

Two other alternative strategies are often used in lattice simulations to determine the three-flavour quark condensate, both being related to the topological properties of QCD on the lattice. The first one consists in studying  the infrared end of the spectrum of the Dirac operator directly, as the accumulation of its eigenvalues around zero provides information on the value of the quark condensate~\cite{Leutwyler:1992yt,DescotesGenon:1999zj}. We will not address this type of determination in the present article, but we notice that a joint JLQCD and TWQCD lattice study of the distribution of low-lying Dirac eigenvalues obtained~\cite{Fukaya:2010na}: $\Sigma(2)/\Sigma(3)=1.30\pm 0.54$, where $\Sigma(N_f)$ denotes the quark condensate in the chiral limit of $N_f$ massless flavours, in agreement with a paramagnetic suppression of chiral order parameters as the number of massless flavours increases.

A second possibility consists in studying the so-called topological susceptibility, which will be the main focus of the present article. One can introduce the winding-number density:
\begin{equation}
\omega=\frac{1}{16\pi^2}{\rm tr}_c\tilde{G}^{a}_{\mu \nu} G_{a}^{\mu \nu}\,,
\end{equation}
where a trace over colour indices is performed, and whose integral $Q= \int dx\ \omega$
can be used to classify the gluonic configurations according to topology. In the case of lattice simulations, the winding number of a gauge configuration can be determined through the choice of appropriate periodic boundary conditions~\cite{Leutwyler:1992yt}. Its conjugate variable in the QCD Lagrangian is the so-called vacuum angle $\theta$, whose value is strongly constrained by the absence of CP violation observed in strong interactions. The value of $\theta$ is experimentally very small ($\theta < 10^{-10}$~\cite{PDG}), which requires either fine tuning or a dynamical explanation (e.g., axions~\cite{Peccei:1977ur,Peccei:1977hh}).

The topological susceptibility defined as  the mean square winding number per unit volume $\chi=\langle Q^2 \rangle/V$ occurs when connecting
an observable computed for a fixed topology (i.e., a fixed winding number) to the same observable in the $\theta$ vacuum at finite volume~\cite{Aoki:2007ka}. Up to $O(1/V^3)$, this relation involves both $\chi$ and the coefficient 
$c_4=[\langle Q^4 \rangle -3 (\langle Q^2 \rangle)^2)]/V$ from the 4-point correlator of winding number densities (we will come back to this second quantity in a forthcoming article~\cite{wip}).
The topological susceptibility is related to the value at zero of the correlation function:
\begin{equation}\label{eq:correlwinding}
\chi(p^2)=-i\int d^4x\, e^{ip\cdot x} \langle 0|T\omega(x)\omega(0)|0\rangle\,.
\end{equation}
This correlation function can be obtained from the generating functional of QCD by performing two derivatives with respect to the local source corresponding to the vacuum angle $\theta(x)$. 
As discussed  in ref.~\cite{Kaiser:2000gs}, this correlation function is too singular in QCD  for the integral to exist, so that eq.~(\ref{eq:correlwinding}) is an ambiguous notion and has to be renormalised.
This problem however does not affect the value of $\chi$ at zero momentum transfer and per unit volume $\chi \equiv \chi(0) /V $, i.e., the topological susceptibility.
Indeed, $\chi$ can be related through Ward identities to correlators requiring no renormalisation:
\begin{equation}\label{eq:wardident}
\chi(0)=-\frac{i}{6}\int dx \langle 0|T\sigma_0(x)\sigma_0(0)|0\rangle-\frac{1}{9}\langle 0|\bar{q} m q|0\rangle\,, \qquad \sigma_0=\sqrt\frac{2}{3} \bar{q}i\gamma_5 m q\,.
\end{equation}
The two terms on the right-hand side of this Ward identity (written in the case of degenerate masses among quarks collected in a flavour multiplet $q$~\cite{Kaiser:2000gs}) shows the connection of the topological susceptibility with the determination of the quark condensate and the propagation of flavour-singlet Goldstone bosons respectively. It is thus of no surprise that this quantity, related to topological properties of QCD, is also exploited to determine the quark condensate on the lattice~\cite{Chiu:2008jq,Bazavov:2010xr,Aoki:2010dy,Chiu:2011dz,Cichy:2011an}.

This is reflected at the level of the low-energy effective theory of QCD, i.e. Chiral Perturbation Theory ($\chi$PT)~\cite{chpt-su2,chpt-su3}.
One builds a generating functional  including the source corresponding to the vacuum angle, $\theta(x)$. The latter is only affected by axial flavour-singlet transformations, and thus connected to that of the (pseudo)scalar source term through:
\begin{equation}\label{eq:axialrotation}
[s(x)+ip(x)=s_0+i p_0,\theta(x)=\theta_0] \leftrightarrow [s(x)+ip(x)=V_R(s_0+ip_0)V_L^\dag,  \theta(x)=\theta_0+\arg\det(V_R^\dag V_L)]\,.
\end{equation}
The source $\theta(x)$ corresponds to the phase of the matrix collecting Goldstone bosons in the building of the low-energy effective theory, yielding
Ward identities connecting correlators with (pseudo)scalar and winding-number densities. Let us note that
a general mass matrix $s_0+i p_0$  can be brought down to a diagonal one with real positive eigenvalues $m_u,m_d,m_s$ through the axial transformation eq.~(\ref{eq:axialrotation}), so that the effective theory depends on $s_0,p_0,\theta_0$ only through $m_u,m_d,m_s$ and $\bar\theta=\theta_0+\arg\det(s_0+ip_0)$. If one of the quark masses is equal to zero, this transformation can also be used to shift the vacuum angle $\theta$ by an arbitrary amount without modifying the theory: the $\chi$PT effective potential becomes independent of $\theta$, and its second derivative, i.e., $\chi$, should vanish in this limit (this can be also seen from the Ward identity eq.~(\ref{eq:wardident}) written for several distinct quark masses).

Eq.~(\ref{eq:axialrotation}) yields a natural connection between $\theta$-related observables, the $U_A(1)$ anomaly and the $\eta'$ meson, connection which can be investigated further by performing a simultaneous expansion in momenta, quark masses and $1/N_c$ and promoting the $\eta'$ meson to a ninth Goldstone boson~\cite{Leutwyler:1992yt,Kaiser:2000gs}. This has led to a detailed study of the topological susceptibility in the pure Yang-Mills theory (see, e.g., refs.~\cite{Durr:2006ky,Giusti:2010qa}). But even without relying on the large-$N_c$ expansion, one can determine the low-energy behaviour of $\chi(p^2)$, and in particular its value at zero, within three-flavour Chiral Perturbation Theory. The leading-order result follows the structure of the vanishing-momentum Ward identity eq.~(\ref{eq:wardident}) closely~\cite{chpt-su3}:
\begin{equation}\label{eq:LOsustop}
\chi(p^2)=-\sum_{P=\pi^0,\eta} \frac{|\langle 0|\omega|P\rangle|^2}{\mo_P^2-p^2}
  +\frac{1}{9} \Sigma(3) (m_u+m_d+m_s)+\frac{1}{6}p^2\tilde{H}_0+{\cal O}(p^4)\,,
\end{equation}
where $\Sigma(3)$ is (the opposite of) the quark condensate in the $N_f=3$ chiral limit~\footnote{In the following, we will denote $\Sigma(N_f)$ and $F(N_f)$ the quark condensate and the pseudoscalar decay constant in the chiral limit where the masses of the lighter $N_f$ quarks vanish. According to refs.~\cite{chpt-su2,chpt-su3}:
\begin{equation}
F_0=F(3), \quad \Sigma_0=\Sigma(3)=F_0^2B_0, \qquad F=F(2), \quad \Sigma=\Sigma(2)=F^2B\,.
\end{equation}}, $\langle 0|\omega|P\rangle$ is the coupling of flavour-singlet Goldstone bosons to the winding number density and $\mo_P^2$ the leading-order (LO) contribution to their masses, while $\tilde{H}_0$ is a high-energy counterterm affecting $\chi'(0)$ but not $\chi(0)$.
Replacing the first term on the left hand-side by its appropriate expression and setting the momentum transfer to zero, one finally gets the  widely used and very simple leading-order expression: 
\begin{equation} \label{eq:LOchi}
\chi = \Sigma(3) \bar m   \,,\qquad 1/\bar{m} \equiv 1/m_u +1/m_d +1/m_s\,,
\end{equation} 
with an interesting non-analytic structure going to zero for vanishing $m_u,m_d$ or $m_s$.
Obviously, an accurate determination of the quark condensate from the topological susceptibility requires to go beyond a LO analysis.
The next-to-leading (NLO) expression for the topological susceptibility was recently computed in ref.~\cite{Mao:2009sy}, by determining the dependence of the $\chi$PT  potential with respect to the vacuum angle (i.e., taking the source term $\theta(x)$ as a constant) and performing the corresponding partial derivatives. 
The LO formula  (\ref{eq:LOchi}) was exploited by the TWQCD collaboration to extract the value of the three-flavour quark condensate~\cite{Chiu:2008jq} from RBC/UKQCD configurations with 2+1 domain-wall fermions~\cite{Allton:2007hx}: $\Sigma(3,2\ {\rm GeV})=[259(6)(9)\ {\rm MeV}]^3$ (in the $\overline{MS}$ scheme). Another study was performed by the RBC and UKQCD collaborations  based on data at larger volumes using the NLO formula~\cite{Aoki:2010dy}, but it was not fully exploited as only a consistency check with the dependence on $m_{u,d}$ was performed and the three-flavour quark condensate was not determined by this method.

In this context, it is interesting to consider the topological susceptibility as a way to extract the quark condensate, taking into account the possibility that it is not the order parameter triggering three-flavour chiral symmetry breaking. Indeed various analyses of lattice results~\cite{Colangelo:2010et,Allton:2008pn,Boyle:2007qe,Boyle:2010bh,Aoki:2008sm,Bernard:2007ps, Bazavov:2009tw} suggest an overall good agreement between lattice simulations and $\chi$PT concerning chiral series obtained as an expansion in $m_u$ and $m_d$ only ($N_f=2$ $\chi$PT), but significant disagreements concerning the chiral expansions in powers of $m_u, m_d, m_s$ ($N_f=3$ $\chi$PT). In some cases, small values of the $N_f=3$ quark condensate and pseudoscalar decay constant fail to saturate the LO chiral expansions. In other cases, the convergence of the three-flavour chiral series exhibits sometimes unexpected features, as small next-to-leading (NLO) contributions and large next-to-next-to-leading (NNLO) ones, or numerical cancellations between LO and NLO terms.

A scenario where such  features could occur was discussed in a series of articles~\cite{DescotesGenon:1999uh,DescotesGenon:2000di,DescotesGenon:2000ct,DescotesGenon:2002yv, DescotesGenon:2003cg,DescotesGenon:2004iu,DescotesGenon:2007ta,Bernard:2010ex}: significant vacuum fluctuations of $s\bar{s}$ pairs, hinted at by the violation of the Zweig rule in the scalar sector, could lead to a
paramagnetic suppression of LO chiral order parameters (quark condensate and pseudoscalar decay constant). In the $N_f=3$ chiral expansions of observables such as decay constants, masses, form factors\ldots these fluctuations would decrease the size of LO contributions and enhance NLO contributions numerically 
 (see also refs.~\cite{Kolesar:2008jr,Kolesar:2008fu} for a discussion of the $\eta$ dynamics in the same framework).
This would require a particular treatment of chiral expansions, going under the name of
Resummed Chiral Perturbation Theory (Re$\chi$PT). It resums higher-order
contributions in chiral series from the $L_4$ and $L_6$ NLO low-energy constants encoding
the effect of $s\bar{s}$ pairs on the structure of the chiral 
vacuum and may induce a significant $m_s$-dependence in the pattern of chiral 
symmetry breaking. It is compatible with the usual 
treatment of chiral series in the limit where the latter are saturated by their LO term, but it allows for a consistent treatment of 
the series even if there is a significant competition of LO and NLO 
contributions for some of the observables.

In ref.~\cite{Bernard:2010ex}, we illustrated our procedure for energy-dependent quantities, namely form factors, such as the electromagnetic form factors of pions and kaons, as well as those for $K_{\ell 3}$ decays. We used the same framework to deal with masses, decay constants and form factors simulated on the lattice for 2+1 dynamical flavours (in the unitary limit where valence and sea quarks have the same masses). 
This allowed us to extract information on the pattern of three-flavour chiral symmetry breaking from lattice data on light-meson spectrum and form factors, with a significant paramagnetic suppression of the quark condensate and the pseudoscalar decay  constant in the three-flavour chiral limit.
In the present article, we aim at confronting the possibility of significant paramagnetic suppression of the quark condensate with  lattice data gathered on the topological susceptibility.
In Sec.~3, we reconsider the derivation of the
NLO expression for the topological susceptibility in the case of a suppression of the LO $N_f=3$ chiral
order parameters. We also discuss the implications of this formula for the determination of the quark
condensate. In Sec.~4, we illustrate our approach by fitting results obtained by the RBC/UKQCD data on the topological susceptibility as well as the pseudoscalar meson spectrum. An important issue consists in the determination of the lattice spacings and the related discretisation errors, which we discuss in Sec. 5 before concluding.

\section{Resummed Chiral Perturbation Theory and $\eta$ observables}\label{sec:rechpt}

We have already discussed the salient features of the resummed $\chi$PT framework~\cite{Bernard:2010ex}. Let us briefly recall the basic ideas keeping in mind that no specific hypothesis is  made upon the relative size of LO and NLO.

\begin{itemize}
\item First,  a subset of observables  is chosen for which a ``good`` overall convergence is assumed, i.e. the sum of LO and NLO terms
 is large compared to the remaining higher-order (HO) terms of the series (in relation with the correlators described by 
$\chi$PT): for example, the  squared pion and kaon decay constants,
 $F_{\pi}^2$, $F_{K}^2$ (related to the two-point axial correlator $\langle A^{\mu} A^{\nu} \rangle$), the squared masses, $F_{\pi}^2M_{\pi}^2$, $F_{K}^2M_{K}^2$ ($\langle \partial_{\mu}A^{\mu}  \partial_{\nu}A^{\nu} \rangle$)  and the pion electromagnetic form factor,  $F_{\pi}F_V^{\pi}$
 ($\langle  A^{\nu} V^{\mu} A^{\sigma} \rangle$). Their chiral expansion is performed in terms of $\chi$PT Lagrangian parameters up to NLO, \textit{keeping track} of the HO which form the (small) remainders of the series. 

\item Furthermore, three fundamental LO quantities 
\begin{equation}\label{eq:LOLECs}
X(3)=\frac{2m\Sigma(3)}{F_\pi^2M_\pi^2}\,,\qquad Z(3)=\frac{F^2(3)}{F_\pi^2}\,,\qquad r=\frac{m_s}{m}\,,
\end{equation}
as well as HO remainders, are kept free.
The first two quantities in 
eq.~(\ref{eq:LOLECs}) are of particular relevance, since they express 
two main order parameters of $N_f=3$ chiral symmetry breaking, the quark 
condensate and the pseudoscalar decay constant, in physical units. They also 
assess the saturation of the chiral expansion of $F_\pi^2M_\pi^2$ and 
$F_\pi^2$ by their leading order. 
One of them can be traded for:
\begin{equation}
Y(3)=\frac{X(3)}{Z(3)}=\frac{2m B_0}{M_\pi^2}\,.
\end{equation}
The ratio $r$
measures the relative size of the quark masses in a framework where the 
strange quark is supposed to play a peculiar role in the chiral structure of 
QCD vacuum. 

\item Finally, the parameters of the Lagrangian which appear at NLO, the so-called low-energy constants (LEC) $L_{i}$, are expressed as functions of $X(3)$, $Z(3)$, $r$, some conveniently chosen observables and the remainders. The relations thus obtained can be inserted into  the chiral expansions of other observables,
introducing additional remainders in an appropriate way. 
\end{itemize}

The important point is \textit{not to systematically trade} leading-order terms for physical ones (for example $2m B_{0}$ is not traded for $M_{\pi}^{2}$), as  usually done, but to perform the replacement only when physical arguments indicate that the convergence of the series will be improved, and thus higher-order remainders will see their size reduced. For example, the nonanalytic structure imposed by unitarity should be located at the physical poles, thresholds\ldots We will not further comment on this question here since we only will encounter tadpoles contributions  in the following. A thorough discussion of an implementation of exact perturbative unitarity and exact renormalization scale independence and its illustration for $\pi \eta$ scattering can be found in ref.~\cite{Kolesar:2008jr}~\footnote{Let us mention that there are slight differences between refs.~\cite{Kolesar:2008jr} and \cite{Bernard:2010ex} concerning the treatment of the unitarity part, which affect the corresponding pieces in the form factors and scattering amplitudes, but not the observables considered here.}.

As indicated by eqs.~(\ref{eq:wardident}) and (\ref{eq:LOsustop}), the propagation and coupling of the $\pi^0$ and $\eta$ mesons will play a particular role for the analysis of the topological susceptibility.
We will thus include a further element that was not discussed in detail in ref.~\cite{Bernard:2010ex}, but presented in ref.~\cite{DescotesGenon:2003cg}, i.e., the decay constant and mass of the $\eta$-meson which features as a propagating mode of the correlator $\chi$. In the physical case and in the isospin limit, we can write down relationships for the $\eta$ observables:
\begin{eqnarray}\label{etadecay}
F_\eta^2 &=& F_\pi^2 Z(3)
  +8(r+2)Y(3)M_\pi^2 \Delta L_4
  +\frac{8}{3}(2r+1)Y(3)M_\pi^2 \Delta L_5\\
&&  +\frac{Y(3)M_\pi^2}{48\pi^2}\left[(2r+1)\log\frac{\mo_\eta^2}{\mo_K^2}-\log\frac{\mo_K^2}{\mo_\pi^2}\right]+
 F_\eta^2e_\eta\,, \nonumber\\
\label{etamass}
F_\eta^2M_\eta^2 &=& \frac{2r+1}{3} F_\pi^2M_\pi^2 X(3)
  +\frac{16}{3}(2r+1)(r+2)Y^2(3)M_\pi^4 \Delta L_6\\
&&
  +\frac{16}{3}(2r^2+1) Y^2(3)M_\pi^4 \Delta L_8
  +\frac{32}{3}(r-1)^2 Y^2(3)M_\pi^4 L_7+ F_\eta^2M_\eta^2d_\eta\,. \nonumber
\end{eqnarray}
$\mo_P^2$ denotes the LO contribution to the mass of the pseudoscalar meson $P=\pi,K,\eta$.
$\Delta L_i=L_i^r-\hat{L_i}^r$ are scale-independent combinations of the NLO low-energy constants $L_i$ and chiral logarithms $\hat{L}_i^r$. Using NLO chiral series with expected good convergence properties, these combinations can be reexpressed in terms of pseudoscalar masses, decay constants, LO parameters and HO remainders, as discussed at length in ref.~\cite{Bernard:2010ex} and briefly recalled in App.~\ref{app:lec}.
The remainders $d_\eta,e_\eta$ are assumed to be small, with the estimate for the order of magnitude
$d_\eta,e_\eta=O(m_s^2)=O(10\%)$ [in practice, we will follow the dimensional estimates described in App.~B in ref.~\cite{Bernard:2010ex}].

In ref.~\cite{Bernard:2010ex}, we used the relationships between $L_{4,5}$ and $F_\pi^2,F_K^2$, as well as those between $L_{6,8}$ and $F_\pi^2M_\pi^2,F_K^2 M_K^2$ to express the NLO low-energy constants (LECs) in terms of measured quantities, HO remainders as well as the three LO parameters eq.~(\ref{eq:LOLECs}). Following the same approach,
we can invert eq.~(\ref{etamass}) to reexpress the low-energy constant $L_7$ in terms of masses and decay constants:
\begin{equation}\label{eq:L7}
[Y(3)]^2L_7 = \frac{1}{32(r-1)^2}\frac{F_\pi^2}{M_\pi^2}
 \Bigg[\frac{3F_\eta^2 M_\eta^2-4F_K^2M_K^2+F_\pi^2M_\pi^2}{F_\pi^2M_\pi^2}-d_{GO}-(r-1)^2[\epsilon(r)+d']
 \Bigg]\,,
\end{equation}
where $F_\eta$ is not measured, but can be computed using eq.~(\ref{etadecay}). We have introduced
the Gell-Mann--Okubo difference of masses and the equivalent for HO remainders:
\begin{equation}
\Delta_{GO}=\frac{3 F^2_{\eta} M^2_{\eta}-4 F^2_K M^2_K + F^2_{\pi} M^2_{\pi}}{F^2_{\pi} M^2_{\pi}}\,,\qquad  
d_{GO}=\frac{3F_\eta^2M_\eta^2}{F_\pi^2M_\pi^2}d_\eta-
  \frac{4F_K^2M_K^2}{F_\pi^2M_\pi^2}d_K+d_\pi\,,
\end{equation}
and the function of $r$:
\begin{equation} \label{funcr}
\epsilon(r) = 2\frac{r_2-r}{r^2-1}, \qquad
r_2= 2\left(\frac{F_KM_K}{F_{\pi}M_{\pi}}\right)^2 -1\sim 36\,,
\end{equation}
so that $\epsilon(r_2)=0$, and $\epsilon(r_1)=1$ with $r_1=2 (F_KM_K)/(F_\pi M_\pi)-1\simeq 8$.

In the following, we will consider simulations with 2+1 dynamical flavours $(\tilde{m},\tilde{m},\tilde{m}_s)$ and denote $\tilde{X}$ the values obtained from lattice simulations (and $X$ the corresponding value for physical quark masses). We introduce the ratios:
\begin{equation}\label{eq:pq}
p=\frac{\tilde{m}_s}{m_s} \,,\qquad q=\frac{\tilde{m}}{\tilde{m}_s}\,,
\end{equation}
in addition to the ratio of physical quark masses $r$ and the chiral parameters arising in the LO Lagrangian in eq.~(\ref{eq:LOLECs}). This yields the following expressions for the $\eta$ mass and decay constant for lattice simulations:
\begin{eqnarray}\label{etamasslatt}
\tilde{F}_\eta^2 &=& F_\pi^2 Z(3)
  +8pqr\left(\frac{1}{q}+2\right)Y(3)M_\pi^2 L_4^r(\mu)
  +\frac{8}{3}pqr\left(\frac{2}{q}+1\right)Y(3)M_\pi^2  L_5^r(\mu)\\
&&  -\frac{M_\pi^2}{32\pi^2}pqrY(3)\log\frac{\tilde\mo_K^2}{\mu}+
 \tilde{F}_\eta^2\tilde{e}_\eta\,, \nonumber\\
\label{etadecaylatt}
\tilde{F}_\eta^2\tilde{M}_\eta^2 &=& \frac{pqr}{3}\left(\frac{2}{q}+1\right) F_\pi^2M_\pi^2 X(3)
  +\frac{16}{3}(pqr)^2\left(\frac{2}{q}+1\right)\left(\frac{1}{q}+2\right)Y^2(3)M_\pi^4 L_6^r(\mu)\\ \nonumber
&&
  +\frac{32}{3}(pqr)^2\left(\frac{1}{q}-1\right)^2 Y^2(3)M_\pi^4 L_7
  +\frac{16}{3}\left(\frac{2}{q^2}+1\right)(pqr)^2 Y^2(3)M_\pi^4 L_8^r(\mu)\\ \nonumber
&& 
  -\frac{M_\pi^4}{32\pi^2}[pqrY(3)]^2
  \Bigg(\frac{1}{3}\left(\frac{2}{q}+1\right)
  \left[2\left(\frac{1}{q}+1\right)\log\frac{\tilde\mo_K^2}{\mu^2}
       +\frac{4}{9}\left(\frac{2}{q}+1\right)\log\frac{\tilde\mo_\eta^2}{\mu^2}
      \right]\\ \nonumber
&&   +\left[\log\frac{\tilde\mo_\pi^2}{\mu^2}
       -\frac{1}{3}\left(\frac{1}{q}+1\right)\log\frac{\tilde\mo_K^2}{\mu^2}
       -\frac{1}{9}\left(\frac{2}{q}+1\right)\log\frac{\tilde\mo_\eta^2}{\mu^2}
      \right]
 \Bigg)+\tilde{F}_\eta^2\tilde{M}_\eta^2\tilde{d}_\eta\,. \nonumber
\end{eqnarray}
where $L_7$ is given by eq.~(\ref{eq:L7}), and equivalent expressions for $L_{4,5,6,8}^r$ can be found in ref.~\cite{Bernard:2010ex} in terms of $r,X(3),Z(3)$, HO remainders, masses and decay constants (they are recalled in App.~\ref{app:lec}). 
The HO remainders $\tilde{d}_\eta, \tilde{e}_\eta$ are supposed to scale like $\tilde{m}_s^2$ for simulations where the simulated strange quark is much heavier than the other ones. The LO contributions to the masses of the simulated pseudoscalar mesons read:
\begin{equation}
\tilde\mo_\pi^2=pqr M_\pi^2 Y(3)\,, \qquad \tilde\mo_K^2=\frac{pqr}{2}\left(\frac{1}{q}+1\right) M_\pi^2Y(3)\,,
 \qquad \tilde\mo_\eta^2=\frac{pqr}{3}\left(\frac{2}{q}+1\right)M_\pi^2 Y(3)\,.
\label{eq:kring}
\end{equation}

\section{The topological susceptibility at next-to-leading order}

\subsection{Diagrammatic analysis}

As discussed in the introduction, the topological susceptibility can be seen as the value at zero of the two-point 
winding number density correlator, or as the averaged winding number density. 
It can be computed in the effective theory by considering the whole 
correlator~(as was done in ref.~\cite{chpt-su3} to obtain the leading-order expression) or by considering only the effective potential for constant sources and determining its dependence on the vacuum angle $\theta$~(like the derivation of the NLO expression in ref.~\cite{Mao:2009sy}). In the latter case, we have performed the calculation keeping all orders in strong isospin-breaking (contrary to ref.~\cite{Mao:2009sy}) and obtained in the $\chi$PT framework~\cite{chpt-su3}:
\begin{eqnarray}\label{eq:chinopoleisobreak}
&&\chi^{\rm eff.\ pot.}=
\bar{m}F_0^2 B_0 +32\bar{m}(m_u+m_d+m_s)B_0^2L_6^r(\mu)+96\bar{m}^2B_0^2(3L_7+L_8^r(\mu))\\
&& \qquad-\frac{\bar{m}^2 B_0}{32\pi^2}\times \Bigg[\sum_{i\neq j}\frac{(m_i+m_j)^2B_0}{m_i m_j}\log\frac{B_0(m_i+m_j)}{\mu^2}\nonumber\\
&&\qquad\qquad+\left(\frac{m_u+m_d}{m_um_d}+\frac{2\sin\epsilon\cos\epsilon}{\sqrt{3}}\frac{m_d-m_u}{m_u m_d}
 +\frac{2}{3}\sin^2\epsilon \frac{2m_um_d-m_s(m_u+m_d)}{m_um_dm_s}
\right)\mo_{\pi^0}^2\log\frac{\mo_{\pi^0}^2}{\mu^2}\nonumber\\
&&\qquad\qquad+\Big( \frac{4m_um_d+m_s(m_u+m_d)}{3m_um_dm_s}-\frac{2\sin\epsilon\cos\epsilon}{\sqrt{3}}\frac{m_d-m_u}{m_u m_d}\nonumber\\
&&\qquad\qquad\qquad\qquad\qquad\qquad\qquad\qquad\qquad\qquad
 -\frac{2}{3}\sin^2\epsilon \frac{2m_um_d-m_s(m_u+m_d)}{m_um_dm_s}
\Big)\mo_{\eta}^2\log\frac{\mo_{\eta}^2}{\mu^2}
 \Bigg]\nonumber\,,
\end{eqnarray}
where $\epsilon$ is the LO $\pi^0\eta$ mixing angle, defined as:
\begin{equation}
\tan 2\epsilon=\frac{\sqrt{3}}{2}\frac{m_d-m_u}{m_s-m}\,, \qquad m=(m_u+m_d)/2\,.
\end{equation}
This expression is indeed independent of the renormalisation scale $\mu$ 
and agrees with ref.~\cite{Mao:2009sy} in the isospin limit, as expected. 
However, this result as it stands does not exploit the Ward identity eq.~(\ref{eq:wardident}),
which suggests strongly that one should identify the $\eta,\pi^0$ poles 
in the chiral expansion of the correlator.

\begin{figure}[t!]
\begin{center}
\includegraphics[width=13cm,angle=0]{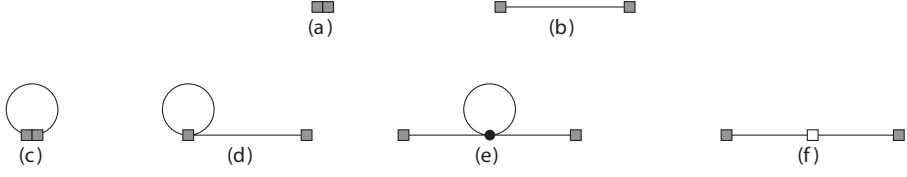}
\end{center}
\caption{\small\it  Diagrams involved in the evaluation of the correlator $\chi$. The winding number densities $\omega$ are represented by the gray boxes, the propagating $\eta$ and $\pi^0$ mesons by the solid lines attached to the sources, whereas the loops contain any of the eight Goldstone bosons. The black dot in diagram (e) corresponds to the usual four-point vertex, and the white square in diagram (f) to the NLO low-energy counterterms.}\label{fig:diags}

\end{figure}

It is straightforward to analyse the correlator $\chi$, eq.~(\ref{eq:correlwinding}), diagrammatically in NLO $\chi$PT to identify the parts corresponding to propagation, as shown in Fig.~\ref{fig:diags}. The LO expression eq.~(\ref{eq:LOsustop}) comes from the local terms in diagram (a) as well as the propagation part in diagram (b). At NLO, the local term gets contributions from diagram (c),
the coupling of the flavour-singlet mesons to the winding number density  receives 
corrections from diagram (d) and the propagation of the mesons is corrected by diagrams (e) and (f), which are partly absorbed by the redefinition of the pole position, yielding: 
\begin{equation}
\chi(q^2)=\sum_{P=\pi^0,\eta} \frac{R_P}{M_P^2-p^2}+S(q^2)\,,
\end{equation}
where $R_P\equiv|\langle 0|\omega|P\rangle|^2$ is the relevant coupling to Goldstone bosons,
and $S$ is analytic up to the next thresholds (the first singularities being the cut from $s\geq 9M_\pi^2$ and pole
at $M_{\eta'}^2$). 
For the topological susceptibility in the isospin limit and at next-to-leading order, one obtains  the separation:
\begin{equation}\label{eq:etapolegen}
\chi^{\rm pole}=\frac{R}{M_\eta^2}+S\,, \qquad R\equiv R_\eta=|\langle 0|\omega|\eta\rangle|^2\,, \qquad S\equiv S(0)\,.
\end{equation}
The $\pi^0$ meson does not couple to the winding number density in this limit, due to $G$-parity. Up to next-to-leading order the pole residue and the analytical piece read:
\begin{eqnarray}\label{eq:RNLO}
\hspace{-0.7cm} R&=&
-\frac{4}{27}B_{0}^2F_{0}^2(m-m_s)^2\left[1+\frac{1}{F_{0}^2}\left(\frac{\mo_K^2}{16\pi^2}\log\frac{\mo_K^2}{\mu^2}-16 B_{0}(L_4^r(\mu)(2m+m_s)+\frac{L_5^r(\mu)}{3}(m+2m_s))\right)\right]\nonumber\\
&&+\frac{4}{27}B_{0}^{2}\frac{(m-m_s)}{16\pi^2}\left[3 m \mo_\pi^2 \log\frac{\mo_\pi^2}{\mu^2}- 2m_s \mo_K^2 \log\frac{\mo_K^2}{\mu^2} + \frac{1}{3}(m-4m_s)\mo_\eta^2 \log\frac{\mo_\eta^2}{\mu^2}   \right]\nonumber\\
&& -\frac{256}{27}B_{0}^{3}(m-m_s)\left[(L_6^r(\mu)+L_7^r(\mu))(m-m_s)(2m + m_s)+L_8^r(\mu)(m^2-m_s^2) \right]\,,
\end{eqnarray}
\begin{eqnarray}
\hspace{-0.7cm} S&=&\frac{1}{9}B_0F_{0}^2(2m + m_s)\nonumber\\
&&-\frac{B_{0}}{144\pi^2}\left[3m \mo_\pi^2 \log\frac{\mo_\pi^2}{\mu^2} + 2(m+m_s)\mo_K^2 \log\frac{\mo_K^2}{\mu^2}  + \frac{1}{3}(2 m_s +m)\mo_\eta^2 \log\frac{\mo_\eta^2}{\mu^2}  \right]\nonumber\\
&&+ \frac{32}{9}B_{0}^2\left[ (2m+m_s )^2(L_6^r(\mu)+L_7^r(\mu))+(2m^2+m_s^2)L_8^r(\mu)\right]\,,
\label{eq:SNLO}
\end{eqnarray}
It is easy to check that eq.~(\ref{eq:etapolegen}) coincides with eq.~(\ref{eq:chinopoleisobreak}) in the isospin limit, up to higher-order terms in the chiral expansion.

\subsection{Resummed expression (no $\eta$ pole)} \label{sec:resumnoeta}

Like in the case of $F_\pi^2M_\pi^2$, the three-flavour quark condensate arises at LO in eq.~(\ref{eq:chinopoleisobreak}), whereas NLO involves the low-energy constant $L_6$, related to the violation of the Zweig rule in the scalar sector and enhanced by a factor of $m_s$. This expression (or even its truncation at LO only) is used to determine the quark condensate, assuming the smallness of NLO corrections.
In refs.~\cite{DescotesGenon:1999uh,DescotesGenon:2000di,DescotesGenon:2000ct,DescotesGenon:2002yv, DescotesGenon:2003cg,DescotesGenon:2004iu,DescotesGenon:2007ta,Bernard:2010ex}, we have argued that the pattern of $N_f=3$ chiral symmetry breaking could be affected significantly by vacuum fluctuations of $s\bar{s}$ pairs, leading to the suppression of the quark condensate, the enhancement of $L_6$ (and  $L_4$) and finally a numerical competition between LO and NLO
contributions in $N_f=3$ chiral expansions. Such a problem would occur for eq.~(\ref{eq:chinopoleisobreak}), which explains why we want to analyse the topological susceptibility in the Re$\chi$PT framework.

We consider now a lattice simulation with dynamical quarks of unphysical masses $(\tilde{m},\tilde{m},\tilde{m}_s)$. If we do not isolate the contribution from the $\eta$ pole and simply reexpress eq.~(\ref{eq:chinopoleisobreak}) using eqs.~(\ref{eq:LOLECs}) and (\ref{eq:pq}), we obtain:
\begin{eqnarray}\label{eq:chinopole}
&&\!\!\!\!\tilde\chi^{\rm no\ pole}=\frac{F_\pi^2M_\pi^2}{2}\frac{pqr}{q+2}\Bigg[X(3)+ 16[Y(3)]^2 \frac{M_\pi^2}{F_\pi^2} pr \left[L_6^r(\mu)(2q+1)+3(3L_7+L_8^r(\mu)) \frac{q}{q+2}\right]\\\nonumber
&& -
   \frac{1}{16\pi^2}[Y(3)]^2 \frac{M_\pi^2}{F_\pi^2} pr
   \Bigg[
       \frac{3q}{q+2}\log\frac{\tilde\mo_\pi^2}{\mu^2}+\frac{(1+q)^2}{q+2}\log\frac{\tilde\mo_K^2}{\mu^2}+\frac{2q+1}{9}\log\frac{\tilde\mo_\eta^2}{\mu^2}\Bigg]\Bigg]+ \tilde\chi^{\rm no\ pole}\tilde{d}_{\chi^{\rm no\ pole}}\,, \qquad
\end{eqnarray}
where $d_\chi$ denotes the remainder collecting higher-order contributions.
If we replace $L_{6,7,8}$ by their expressions in terms of LO quantities and physical observables, eq.~(\ref{eq:L7}) and App.~\ref{app:lec}, and if we consider a simulation at the physical point ($q=1/r,p=1$), the topological susceptibility boils down to:
\begin{eqnarray}\label{eq:chinopolephysical}
\chi^{\rm no\ pole}&=&\frac{F_\pi^2M_\pi^2}{2}\frac{r}{2r+1}
 \Bigg[
   1-\epsilon(r)\frac{7r+2}{2r(2r+1)} 
+\frac{9}{2}\frac{r}{(r-1)^2(r+2)}\Delta_{GO}\\
&& -\frac{1}{16\pi^2}[Y(3)]^2 \frac{M_\pi^2}{F_\pi^2} r
   \left[
       \frac{3}{2r+1}\log\frac{\mo_\pi^2}{\mu^2}+\frac{(1+r)^2}{r(2r+1)}
\log\frac{\mo_K^2}{\mu^2}+\frac{r+2}{9r}\log\frac{\mo_\eta^2}{\mu^2}\right]\nonumber\\
&& + 4[Y(3)]^2 \frac{M_\pi^2}{F_\pi^2} \left[\hat{L}_6^r(\mu)(r+2)+3\hat{L}_8^r(\mu) \frac{r}{2r+1}\right]\Bigg]+ 
\ldots\nonumber
\end{eqnarray}
where 
the ellipsis denotes higher-order remainders, and 
the combinations of chiral logarithms $\hat{L}_i^r$, which depend on $Y(3)$ and quark masses, are given in App.~\ref{app:lec}.
Eq.~(\ref{eq:chinopolephysical}) shows that in the physical case, the  topological susceptibility has no sensitivity on the three-flavour condensate $X(3)$. Indeed, 
up to $1/r$-suppressed corrections (which are small since $r\geq 15$~\cite{DescotesGenon:2007ta,Batley:2007zz,:2009nv}), eq.~(\ref{eq:chinopole}) combines 
 the (LO) three-flavour condensate and the (NLO)  low-energy constant $L_6$ in the same way as $F_\pi^2M_\pi^2$, or equivalently the two-flavour condensate $X(2)$:
\begin{equation}
X(2)=\lim_{m\to 0}\frac{F_\pi^2M_\pi^2}{2m}
  =F_\pi^2M_\pi^2 X(3)+16r[Y(3)]^2 M_\pi^4 L_6^r(\mu) + \ldots
\end{equation}
which is determined from the study of $\pi\pi$ scattering through $K_{\ell 4}$ decays~\cite{Colangelo:2001sp,DescotesGenon:2001tn,Batley:2007zz,:2009nv,Colangelo:2008sm,DescotesGenon:2012gv,wip2}. As was discussed in ref.~\cite{DescotesGenon:1999uh}, the two-flavour condensate can be significantly different from the three-flavour one,
and its value is strongly correlated to the ratio of quark masses $r$  through
 the three-flavour chiral expansion of $F_\pi^2M_\pi^2$. This feature explains both the absence of $X(3)$ and the $r$-dependence in eq.~(\ref{eq:chinopolephysical}).

\begin{figure}[t!]
\begin{center}
\includegraphics[width=11cm,angle=0]{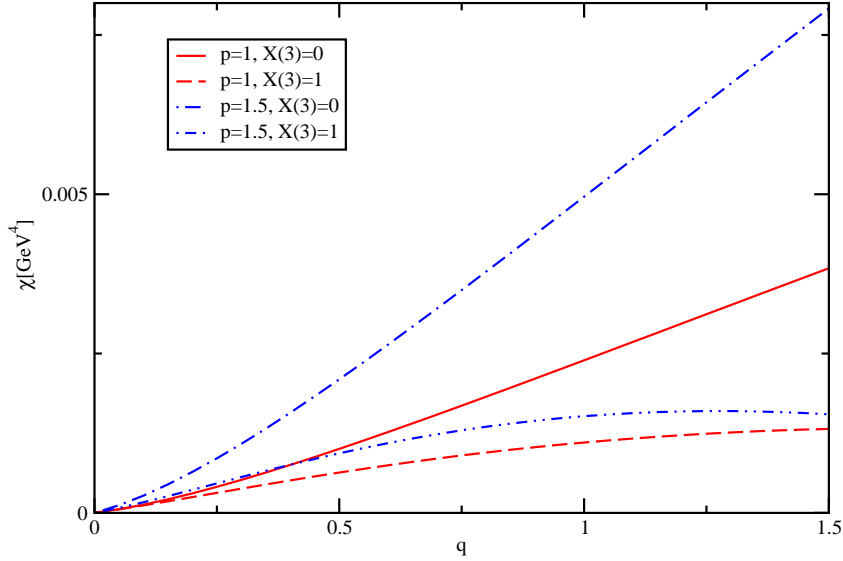}
\end{center}
\caption{\small\it  The topological susceptibility $\tilde\chi^{\rm no\ pole}$ as a function of the ratio of simulated quark masses $q$ for $X(3)=0$ (solid lines) and $X(3)=1$ (dashed lines) for two different values of $p$ (related to  the simulated strange quark mass). For illustration, the remainders are set to be zero, $Y(3)=0.8$, $r=26$, $F_K/F_\pi=1.19$ and $F_\eta=$130 MeV.  \label{figure:results00}}
\end{figure}

Therefore, the topological susceptibility
close to the physical point ($p=1$, $r=1/q$ large) exhibits a rather poor sensitivity to the three-flavour condensate $X(3)$.   In a similar way to pseudoscalar masses and decay constants, simulations aiming at disentangling the pattern of three-flavour chiral symmetry breaking should be performed not only for quark masses with a hierarchy similar to the physical case, but also light (almost) degenerate masses with values between the physical $m_u,m_d$ and $m_s$.
The (unphysical) region where simulations probe $X(3)$ efficiently can be determined by expanding eq.~(\ref{eq:chinopole}) in powers of $1/r$:
\begin{equation}
\tilde\chi^{\rm no\ pole}=\frac{F_\pi^2M_\pi^2}{2}\frac{pqr}{q+2}
\left[X(3)[1-p(2q+1)]+p(2q+1)-\frac{3pq}{2(q+2)}\frac{r_2-r}{r}\right]+\ldots
\end{equation}
where the ellipsis denote chiral logarithms and  HO remainders.  It is clear from this expression that we need $q=O(1)$ rather than $O(1/r)$ to get a reasonable sensitivity to 
$X(3)$: for instance, $q=1$ (i.e., $\tilde{m}=\tilde{m}_s$ yields $[1-3p]X(3)+p[3-(r_2-r)/(2r)]$, so that the 
coefficient of $X(3)$ has a similar size to that of the constant term. 
This is sketched in a more quantitative way in Fig.~\ref{figure:results00} 
which compares the topological susceptibility at $X(3)=0$ and $X(3)=1$ in some illustrative cases~\footnote{For this comparison, we assume that the physical value of $F_\eta$ is known in order to compute the Gell-Mann--Okubo contribution $\Delta_{GO}$ to $L_7$. In the following sections, we will not make this assumption and we will use eq.~(\ref{etadecay}) to determine $F_\eta$.}. The comparison between the two values indicates the sensitivity to the size of the three-flavour quark condensate, which tends to decrease the value of the topological susceptibility. We see that the contributions from $X(3)$ are at least twice as small as the remaining contributions, and they in particular might be hidden in the uncertainties if $q$ is small (e.g., of $O(1/r)$ if the simulated quark masses are tuned to be close to the physical ones).

\begin{figure}
\begin{center}
\includegraphics[width=11cm,angle=0]{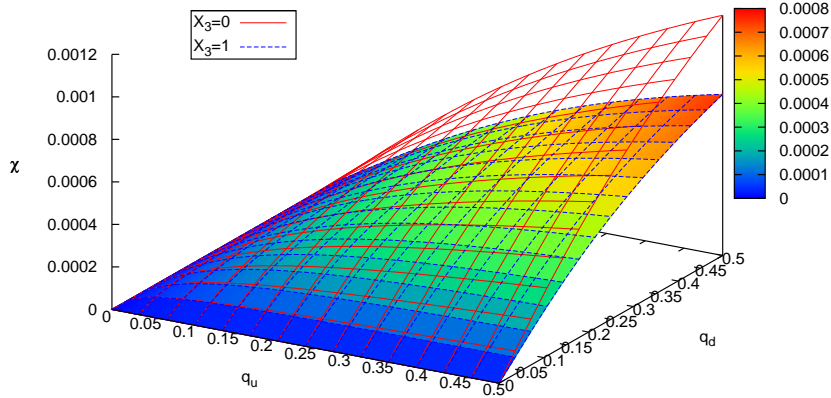}
\end{center}
\caption{\small\it  
The topological susceptibility $\tilde\chi^{\rm no\ pole}$ (in units of $GeV^{4}$) as a function of $q_u= \tilde m_u /\tilde m_s$ and 
$q_d=\tilde m_d /\tilde m_s$ for $X(3)=0$ (solid) and $X_3=1$ (dashed with colour levels). For illustration, the remainders are set to be zero, $p=1$, $Y(3)=0.8$, $r=26$, $F_K/F_\pi=1.19$ and $F_\eta=$130 MeV. 
\label{figure:results01}}
\end{figure}

Another way of escaping the poor sensitivity of the topological susceptibility to the three-flavour quark condensate consists in simulations involving significant isospin breaking, which can be easily implemented in particular for twisted-mass fermion actions~\cite{Baron:2010bv,Cichy:2011an}. 
An expression for the topological susceptibility on the lattice  similar eq.~(\ref{eq:chinopole}) can be written, involving the two quantities 
$q_u= \tilde m_u /\tilde m_s$ and
$q_d=\tilde m_d /\tilde m_s$ instead of $q$: 
\begin{eqnarray}
\tilde\chi^{\rm no\ pole}&=&
\frac{F_\pi^2M_\pi^2}{2}
  \frac{pq_uq_dr}{q_u+q_d+q_uq_d}\\
&& \quad\times  \Bigg[X(3)\left\{1-\frac{pr(1+q_u+q_d)}{r+2}\right\}
   -\epsilon(r)pr\left\{\frac{3q_uq_d}{2(q_u+q_d+q_uq_d)}+\frac{1+q_u+q_d}{r+2}\right\}\nonumber\\
 &&\qquad\qquad
   +\frac{pr(1+q_u+q_d)}{r+2}+\frac{9pq_uq_dr}{2(r-1)^2(q_u+q_d+q_uq_d)}\Delta_{GO}
   \Bigg]+\ldots\nonumber
\end{eqnarray}
where we expressed the NLO LECs $L_{6,7,8}$ using App.~\ref{app:lec} (we have checked that including the $m_d-m_u$ difference in the expressions of App.~\ref{app:lec} would not modify our conclusions). The ellipsis denotes chiral logarithms and HO remainders.
The dependence of $\chi$ on the two quantities $q_u$ 
and $q_d$ is illustrated in Fig.~\ref{figure:results01}. As expected, $\chi(q_u,q_d)=\chi(q_d,q_u)$ 
vanishes for $q_u=0$ or $q_d=0$ since the theory becomes independent of $\theta$ when one of the quark masses vanishes. The value of the topological 
susceptibility increases when one of the two quark masses increases, the other one being kept 
fixed. As clearly seen in Fig.~\ref{figure:results01},
the contribution to the topological susceptibility 
independent of $X(3)$  increases faster than 
the one proportional to $X(3)$, so that a scan through values of $(\tilde{m}_u,\tilde{m}_d)$ at fixed $\tilde{m}_s$ could help to determine unambiguously the contribution (and thus the value) of the three-flavour quark condensate.

\subsection{Resummed expression (identifying the $\eta$ pole)}

Identifying the poles corresponding to the propagation of the pseudoscalar mesons is of particular relevance within Resummed $\chi$PT. In this framework, smaller HO remainders are expected when singularities of the expansion (and in particular poles) 
are correctly located at the physical positions. 
Let us notice that a similar discussion takes place when one considers the two-point correlators of the axial current and/or pseudoscalar density, where
the propagation of a single Goldstone boson occurs also at LO. Indeed, in refs.~\cite{DescotesGenon:1999uh,DescotesGenon:2000di,DescotesGenon:2000ct,DescotesGenon:2002yv,DescotesGenon:2003cg,DescotesGenon:2004iu,DescotesGenon:2007ta,Bernard:2010ex}, we have considered the decay constants and masses as they could be obtained from the NLO expression of $\langle A_\mu A_\nu\rangle$ where the propagators are explicitly expressed in terms of the physical masses, considering either the correlator at 0 (for $F_\pi^2$) or its residue at the pseudoscalar pole (for $F_\pi^2 M_\pi^2$).

Thanks to our diagrammatical analysis of the two-point correlator $\chi$ in $\chi$PT, we can easily identify the contribution from the propagation of the $\eta$ meson in eq.~(\ref{eq:chinopole}), and we obtain the expressions for the pole residue and the analytical piece arising in eq.~(\ref{eq:etapolegen}):
\begin{eqnarray}\label{eq:chipole1}
 \tilde{R}&=&-\frac{(pr(q-1))^2}{27}F_\pi^2M_\pi^4 X(3)Y(3) +\frac{8}{27}(pr)^3(1-q)^2 M_\pi^6Y(3)^3
      \left[(2q+1)L_4^r+\frac{q+2}{3}L_5^r\right]\\
&& -\frac{32}{27}(pr)^3 Y(3)^3 M_\pi^6 [(L_6^r+L_7)(q-1)^2(2q+1)+L_8^r(q^2-1)(q-1)]\nonumber\\
&& +\frac{(pr)^3(q-1)}{864\pi^2} M_\pi^6Y(3)^3
   \left[6q^2\log\frac{\tilde\mo_\pi^2}{\mu^2}-(1+q)^2\log\frac{\tilde\mo_K^2}{\mu^2}
     +\frac{2}{9}(q-4)(q+2)\log\frac{\tilde\mo_\eta^2}{\mu^2}\right]\,,\nonumber
\end{eqnarray} 
\begin{eqnarray} \label{eq:chipole2}
\tilde{S}&=& \frac{pr(2q+1)}{18}F_\pi^2 M_\pi^2 X(3) +\frac{8}{9}(pr)^2Y(3)^2M_\pi^4[(2q+1)^2(L_6^r+L_7)+(2q^2+1)L_8^r]
\\
 &&
-\frac{(pr)^2}{576\pi^2}Y(3)^2M_\pi^4
 \left[6q^2\log\frac{\tilde\mo_\pi^2}{\mu^2}+2(q+1)^2\log\frac{\tilde\mo_K^2}{\mu^2}
  +\frac{2}{9}(2+q)^2\log\frac{\tilde\mo_\eta^2}{\mu^2}\right]\,,\nonumber
  \end{eqnarray}
where the counterterms can be expressed in terms of $r,X(3),Z(3)$ and the remainders. The topological susceptibility then reads:
\begin{equation}\label{eq:chipoleunsub}
\tilde\chi^{\rm pole}=\frac{\tilde{R}}{\tilde{M}_\eta^2}+\tilde{S} + \tilde\chi^{\rm pole}\tilde{d}_{\chi^{\rm pole}} \,.
\end{equation}

At this stage, one has still to discuss the direct remainder $\chi  d_\chi$ which corresponds to higher-order terms, and has been added in eq.~(\ref{eq:chipoleunsub}) as well as in eq.~(\ref{eq:chinopole}). One wants $d_\chi$ to start at NNLO, and thus to be $O(m_q^2)$. 
This expectation can be checked by considering various chiral limits sending $\tilde{m}$ and/or $\tilde{m}_s$ to zero. As already explained previously, the topological susceptibility $\chi$ should vanish in the limit where one of the quark masses goes to zero.
In our framework, this can be translated as $\chi\to 0$ for the three chiral limits:
\begin{itemize}
\item $\tilde{m}_s\to 0$, $\tilde{m}$ fixed: $p\to 0$, $p q$ fixed,
\item $\tilde{m}\to 0$, $\tilde{m}_s$ fixed: $q\to 0$, $p$ fixed,
\item $\tilde{m},\tilde{m}_s\to 0$, $\tilde{m}/\tilde{m}_s$ fixed: $p\to 0$, $q$ fixed.
\end{itemize}
By inspection, one can check that the sum of LO and NLO in eq.~(\ref{eq:chinopole}) does vanish in these three limits, whereas eqs.~(\ref{eq:chipole1})-(\ref{eq:chipole2}) do vanish in these limits only up to 
nonzero higher-order corrections.
Therefore, if we consider the expression for the topological susceptibility not singling out the $\eta$-pole, eq.~(\ref{eq:chinopole}), we can add a higher-order remainder of the form $\tilde\chi^{\rm no\ pole}\tilde{d}_{\chi^{\rm no\ pole}}$ with $\tilde{d}_\chi^{\rm no\ pole}=O(\tilde{m}_q^2)$ that has no singularities when any $\tilde{m}_q\to 0$. In the situations that we consider (one of the masses much larger than the other ones), we expect $\tilde{d}_\chi^{\rm no\ pole}$ to be dominated by $\tilde{m}_s^2$ contributions.

On the other hand, if we single out the $\eta$-pole contributions following eq.~(\ref{eq:chinopole}), we would need $\tilde{d}_\chi^{\rm pole}$ to become singular in the chiral limits considered before, so that $\tilde\chi^{\rm pole}\tilde{d}_{\chi^{\rm pole}}$ does not vanish and cancels the non-vanishing value of eqs.~(\ref{eq:chipole1})-(\ref{eq:chipole2}) coming from higher-order terms. A more satisfying solution consists in 
subtracting these higher-order pieces from the HO remainder, so that $d_\chi$ does not exhibit singularities in any of the chiral limits. In other words, using eqs.~(\ref{eq:chipole1})-(\ref{eq:chipole2}), we write the topological susceptibility as:
\begin{equation}\label{eq:etapole}
C(\tilde{m},\tilde{m}_s)=\frac{\tilde{R}}{\tilde{M}_\eta^2}+\tilde{S}\,,
\qquad \tilde\chi^{\rm pole}=
  C(\tilde{m},\tilde{m}_s)
 -C(0,\tilde{m}_s)
 -C(\tilde{m},0)+\tilde\chi^{\rm pole}\tilde{d}_{\chi^{\rm pole}}\,.
\end{equation}
One can check that the subtracted terms $C(0,\tilde{m}_s)=O(\tilde{m}_s^2)$ and $C(\tilde{m},0)=O(\tilde{m}^2)$ are indeed HO terms. With the definition eq.~(\ref{eq:etapole}), which will be used in the following,
$\tilde{d}_\chi^{\rm pole}$ is regular in the chiral limits described above, and can be considered as $O(\tilde{m}_s^2)$
for the simulations considered here. 

Finally, one should notice that $\tilde{R}$ and $\tilde{S}$ involve again not only the three-flavour condensate $X(3)$ but also NLO LECs, and in particular $L_6$. This dependence is equivalent to
that of eq.~(\ref{eq:chinopole}) up to higher orders in the expansion of the quark masses. It is therefore of no surprise that we find only small numerical differences between eq.~(\ref{eq:chinopole}) and eqs.~(\ref{eq:chipole1})-(\ref{eq:chipole2}) in the range of quark masses of physical interest, and that the previous conclusion concerning the potential of the topological susceptibility to determine the three-flavour quark condensate still applies in this setting.

\begin{figure}[t!]
\begin{center}
\includegraphics[width=11cm,angle=0]{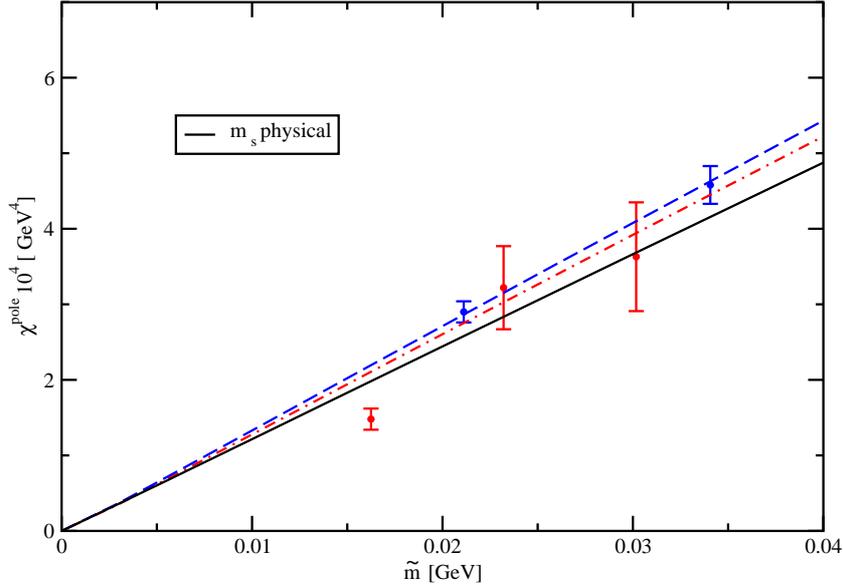}
\end{center}
\caption{\small\it The topological susceptibility $\tilde\chi^{\rm pole}$ as a function of the light quark mass. The black [red] points correspond to 
the $(24)^3\times 64$ volume [$(32)^3\times 64$], using the estimates of lattice spacings quoted in ref.~\cite{Aoki:2010dy}. The dashed curves correspond to the best fit using $\chi^{\rm pole}$ with finite-volume effects, as indicated in the last column of table 1 (removing finite-volume effects would lead to almost identical curves). We have not included the lightest point in our analysis, as discussed in the text. As a reference, we have also indicated the same variation when the strange quark mass is set to its physical value.\label{figure:results0} }
\end{figure}

\subsection{Finite-volume effects}\label{sec:finvol}

Two different lattice artefacts can affect the results of simulations before considering the continuum limit. The first corresponds to the finite size of the lattice spacing, which can be included in principle in the chiral expansion through spurions that depends on the implementation of the fermion action (see ref.~\cite{Sharpe:2006pu} and references therein). For the moment, we will not include this effect as we will consider data with good chiral properties and only $O(a^2)$ artefacts, even though this issue is naturally of interest~\cite{wip}.

A second effect comes from the finiteness of the volume used for lattice simulations, inducing finite-volume modifications of the chiral expansions.
As discussed in ref.~\cite{Luscher:1985dn,Leutwyler:1992yt,Becirevic:2003wk,Colangelo:2003hf,Colangelo:2004xr}, the NLO finite-volume effects for chiral perturbation theory amount to a modification of the chiral logarithms:
\begin{equation}
\frac{\mo_P^2}{16\pi^2}\log\frac{\mo_P^2}{\mu^2}
 \to \frac{1}{2L^3} \sum_{\ell} \frac{1}{\omega_P}=\frac{1}{2L^3}\sigma_P\,,
\end{equation}
where $\ell\in 2\pi/L \times Z^3$ and $\omega_P=\sqrt{\ell^2 + \mo_P^2}$. The summation over (Fourier conjugates of) the three spatial directions comes from the quantization of momenta due to the periodic boundary conditions on the lattice box (making it a torus in practice). We consider simulations where the time component is significantly larger than the spatial components, and consider only the finite-volume effects related to the latter~\cite{Becirevic:2003wk,Colangelo:2003hf,Colangelo:2004xr}. One can make contact
with the infinite-volume logarithm through the function:
\begin{equation}\label{eq:XiP1}
\frac{\sigma_P}{L^3}
  =\frac{\mo_P^2}{8\pi^2}\log\frac{\mo_P^2}{\mu^2}+\Xi_P\,,
\end{equation}
where
$\Xi_P=\xi_{1/2}(L,\mo_P^2)$ was introduced in ref.~\cite{Becirevic:2003wk}. An alternative definition of $\Xi_P$ was proposed in ref.~\cite{DescotesGenon:2004iu} and is rediscussed in App.~\ref{app:XiP}.

We will take into account these corrections in the following analysis for all the quantities  of interest. For masses and decay constants, the corrections can be read directly from the expressions in ref.~\cite{Bernard:2010ex}. For the topological susceptibility, 
we have the following correction from finite-volume effects to the expression without singling out the $\eta$-pole:
\begin{equation}
\chi^{\rm no\ pole}(L)=\chi^{\rm no\ pole}(\infty)-\frac{1}{8}Y(3)M_\pi^2\frac{pqr}{(q+2)^2}
  \left[6\Xi_\pi+4(q+1)\Xi_K +\frac{2}{3}(2q+1)\Xi_\eta\right]\,,
\end{equation}
and when one singles out the $\eta$-pole:
\begin{eqnarray}
\tilde{R}(L)&=&\tilde{R}(\infty) +\frac{(pr)^2(q-1)}{54}Y(3)^2M_\pi^4  \left[3q \Xi_\pi-(1+q) \Xi_K+\frac{1}{3}(q-4)\Xi_\eta\right]\,,\\
\tilde{S}(L)&=&\tilde{S}(\infty)-\frac{(pr)}{72}Y(3)M_\pi^2
 \left[6q\Xi_\pi+4(q+1)\Xi_K+\frac{2}{3}(2+q)\Xi_\eta\right]\,.
 \end{eqnarray}
 We neglect any dependence of the higher-order remainders on finite-volume effects, effectively identifying the HO remainders at finite and infinite volumes.

\section{A first series of fits to lattice data} \label{sec:fittodata}

We want now to exploit the topological susceptibility to improve our determination of the pattern of three-flavour chiral symmetry breaking. We have however seen that the usual setting of lattice simulations is not appropriate, as the topological susceptibility is then driven by the two-flavour condensate. If we want to extract information on the three-flavour pattern, we need either to make simulations away from the physical case (as discussed in Sec.~\ref{sec:resumnoeta}), or to supplement the topological susceptibility with other sources of information.
We will now take this second option, which is also required due to the number of parameters involved in the NLO expression of the topological susceptibility (in particular the parameters of the LO chiral Lagrangian and HO remainders). We will therefore include pseudoscalar masses and decay constants in our considerations and test the compatibility of the topological susceptibility with the pattern of three-flavour chiral symmetry breaking obtained from the latter data.

There are several lattice calculations of the topological susceptibility in the literature, e.g.~refs.~\cite{Chiu:2008jq,Aoki:2010dy}. 
As an illustration of our approach, we will focus on the ones from RBC/UKQCD following our work, ref.~\cite{Bernard:2010ex}, as this collaboration provides all the details (masses, decay constants and topological susceptibility) necessary for our analysis.
Since the publication of ref.~\cite{Bernard:2010ex}, new data from this collaboration have been issued with a new volume and different quark masses, given in ref.~\cite{Aoki:2010dy}, and we will use them as a reference in the coming sections. 
We follow the approach of ref.~\cite{Bernard:2010ex} and perform a fit to the two sets of data in the $24^3\times 64$ ($a^{-1}=1.78(3)$ GeV) and $32^3 \times 64$ volumes ($a^{-1}=2.28(3)$ GeV) with and without including
finite volume effects. In this section, the lattice spacings are fixed to the central values quoted in 
ref.~\cite{Aoki:2010dy} without uncertainty.

The fits include the data collected in App.~\ref{app:data}, i.e. pion and kaon masses and decay constants as well as the topological susceptibility (unfortunately, no lattice data on the $\eta$ meson is available, even though it would improve the analysis of the topological susceptibility in a significant way). We build a $\chi^2$ depending on the following parameters:
\begin{itemize}
\item the three leading-order parameters $r,X(3),Z(3)$,
\item a reference ratio between a simulated strange quark (chosen conventionally to be for the $24^3$ simulations) and the physical strange quark mass,
\item the HO remainders associated with the pion and kaon masses and decay constants (denoted $d,e,d',e'$),
\item the ratio $F_K/F_\pi$ (on the other hand, we set $F_\pi=92.2$ MeV),
\item if the topological susceptibility is included, the corresponding HO remainders for the $\eta$ mass and decay constant ($d_\eta,e_\eta$) as well as the one for the topological susceptibility ($d_\chi$).
\end{itemize}
HO remainders are restricted to remain within a range based on a naive dimensional analysis, as described in ref.~\cite{Bernard:2010ex}.
At this stage, we include no discretisation error effects in our three-flavour chiral expansions, but we will come back to this issue in Sec.~\ref{sec:latticespacing}

As shown in Tab.~\ref{table:results0}, the pattern of $N_f=3$ chiral symmetry breaking with low quark condensate and decay constant observed in that reference is confirmed by this new analysis, whether we include or not the topological susceptibility in our fit. The outcome of the fit is thus mainly driven by the spectrum of pseudoscalar mesons, but the quality of the agreement is not modified by the inclusion of the topological susceptibility, which thus exhibits a good compatibility with the rest of the fit and is consistent with the pattern of chiral symmetry breaking described in ref.~\cite{Bernard:2010ex}.
As in our previous work, $L_4$ and $L_6$ do not show 
any sign of Zweig suppression and the competition between LO and NLO in three-flavour chiral expansions is clearly seen.  We obtain values for the $N_f =2$ chiral order 
parameters in agreement with expectations from two-flavour chiral perturbation theory~\cite{chpt-su2} as well as experimental information on $\pi\pi$ scattering, such as  that from $K_{l4}$ decays~\cite{Colangelo:2001sp,DescotesGenon:2001tn,Batley:2007zz,:2009nv,Colangelo:2008sm,DescotesGenon:2012gv,wip2}. The values of  $\bar  \ell_3$ and  $\bar \ell_4$ given there can be compared to the one obtained from the RBC/UKQCD collaboration, ref.~\cite{Aoki:2010dy}: $\bar 
\ell_3 =2.82 (16), \bar\ell_4 =3.76 (9)$ in infinite-volume $\chi$PT and $\bar \ell_3=2.57 (18), \bar \ell_4=3.83 (9) $ in finite-volume $\chi$PT. One can also recall the value quoted by the Flavour Lattice Averaging Group~\cite{Colangelo:2010et} for $\bar\ell_3=3.2(8)$ (no value was quoted for $\bar\ell_4$ in this reference).
Finally, one notices that singling out the $\eta$-pole or not in the expression of the topological susceptibility does not modify significantly the analysis -- the values obtained with  eqs.~(\ref{eq:chinopole}) and (\ref{eq:chipole1})-(\ref{eq:chipole2}) are very close numerically for the ranges of parameters considered here.

\begin{table}[t!]
$$
\begin{array}{|c||c|c||c|c|c|}
\hline
{\rm Fit\ name} & A1 & A2 & A3 & A4 & A5\\
\hline
\chi &  {\rm No} &  {\rm No} & \chi^{\rm no pole} &  \chi^{\rm pole}&   \chi^{\rm pole}\\
{\rm Finite\ volume} & {\rm No} & {\rm Yes} & {\rm No} & {\rm No} &  {\rm Yes}\\
\hline\hline
\phantom{xx}&&&&&\\[-2.4ex]
r         & 23.0\pm 0.7 & 23.4\pm 0.7 & 23.0 \pm 0.7 & 23.0\pm 0.7&  23.4\pm 0.7\\
X(3)      & 0.33\pm 0.10& 0.34\pm 0.05& 0.33 \pm 0.06& 0.33\pm 0.06& 0.34\pm 0.05\\
Y(3)      & 0.49\pm 0.15& 0.53\pm 0.07& 0.49 \pm 0.08& 0.49\pm 0.08& 0.53\pm 0.07\\
Z(3)      & 0.68\pm 0.03& 0.65\pm 0.03& 0.68 \pm 0.03& 0.68\pm 0.03& 0.65\pm 0.03\\
F_K/F_\pi & 1.17\pm 0.01& 1.17\pm 0.01& 1.17 \pm 0.01& 1.17\pm 0.01& 1.17\pm 0.01\\
{\rm Rem.\ at\ limit}  & d & d & d,e,d_\chi& d,e,d_\chi& d,e_\eta,d_\chi\\
\tilde{m}_{s}(24^3)/m_s 
         & 1.12\pm 0.02& 1.12\pm 0.02&  1.12\pm 0.02& 1.12 \pm 0.02 & 1.12\pm 0.02
         \\
\hline
\phantom{xx}&&&&&\\[-2.4ex]
m_s(2\ {\rm GeV}) [{\rm MeV}] & 98.1\pm 1.7 & 98.4\pm 1.8 & 98.0\pm 1.7 &  98.0\pm 1.7& 98.4\pm 1.8\\
m(2\ {\rm GeV})   [{\rm MeV}] & 4.3\pm 0.1  &  4.2\pm 0.1 & 4.3 \pm 0.1 & 4.3\pm 0.1  & 4.2 \pm 0.1\\
\Sigma_0^{1/3}(2\ {\rm GeV}) [{\rm MeV}] &
186\pm 19& 211\pm 7 & 186\pm 11 & 186\pm 12 & 189\pm 10\\
B_0(2\ {\rm GeV}) [{\rm GeV}] & 1.12\pm 0.34& 1.22\pm 0.17& 1.11\pm 0.21& 1.12\pm 0.21& 1.22\pm 0.17\\
F_0 [{\rm MeV}]               & 75.9\pm 1.3 & 74.4\pm 1.4 & 75.9\pm 1.3 & 75.9\pm 1.3 & 74.5\pm 1.4\\
F_\eta [{\rm MeV}]               & - & - & 117 \pm 16 & 128 \pm 10 & 124\pm 6\\
\chi\cdot 10^4 [{\rm GeV}^4]  & - & - & 0.51\pm 0.01 & 0.49\pm 0.02& 0.50\pm 0.02\\
\tilde{m}_{s}(32^3)/m_s 
         &   1.08 \pm 0.02      & 1.09 \pm 0.02   & 1.07 \pm 0.02 & 1.08\pm 0.02& 1.07\pm 0.02 \\
\hline
\phantom{xx}&&&&&\\[-2.4ex]
 L_4(\mu)\cdot 10^3  & 1.12\pm 0.46& 0.65\pm 0.45& 1.13\pm 0.30& 1.12\pm 0.30 & 0.60\pm 0.43\\
 L_5(\mu)\cdot 10^3  & 2.13\pm 0.78& 2.05\pm 0.40& 2.14\pm 0.53& 2.13\pm 0.53 & 2.04\pm 0.39\\
 L_6(\mu)\cdot 10^3  & 3.00\pm 3.05& 2.55\pm 0.89& 3.13\pm 1.44& 3.10\pm 1.42 & 2.52\pm 0.88\\
 L_7(\mu)\cdot 10^3  & -          & -      & -2.60\pm 1.02& -1.82\pm 0.47 & -1.70\pm 0.40\\
 L_8(\mu)\cdot 10^3  & 4.12\pm 2.74& 3.39\pm 1.13& 4.17\pm 1.79& 4.12\pm 1.77& 3.35\pm 1.11\\[0.15ex]
 \hline
 \phantom{xxx}&&&&&\\[-2.4ex]
X(2)  & 0.90\pm 0.01& 0.90\pm 0.01& 0.90\pm 0.01& 0.90\pm 0.01& 0.90\pm 0.01\\
Y(2)  & 0.99\pm 0.01& 1.00\pm 0.01& 0.99\pm 0.01& 0.99\pm 0.01& 1.00\pm 0.01\\
Z(2)  & 0.91\pm 0.01& 0.91\pm 0.01& 0.91\pm 0.01& 0.91\pm 0.01& 0.90\pm 0.01\\
\Sigma^{1/3}(2\ {\rm GeV}) [{\rm MeV}] &
260\pm 3& 261 \pm 3 & 260\pm 2 & 260\pm 3 & 261\pm 2\\
B (2\ {\rm GeV}) [{\rm GeV}] & 2.26\pm 0.07& 2.31\pm 0.06& 2.26\pm 0.07& 2.26 \pm 0.07 & 2.31\pm 0.06\\
F [{\rm MeV}]                & 88.0\pm 0.4 & 87.8\pm 0.2& 88.1\pm 0.2& 88.0\pm 0.22    & 87.8\pm 0.14\\
\bar\ell_3      & -1.6\pm 1.4& -0.5\pm 0.9& -1.6\pm 1.2& -1.6\pm 1.2 & -0.5\pm 0.9\\
\bar\ell_4      & 3.0\pm 0.3 & 3.2\pm 0.1& 3.0 \pm 0.1& 3.0\pm 0.2  & 3.2\pm 0.1\\[0.15ex]
 \hline
 \phantom{xxx}&&&&&\\[-2.4ex]
\Sigma/\Sigma_0=\Sigma(2)/\Sigma(3) & 2.72\pm 0.78& 2.64\pm 0.35& 2.74 \pm 0.46& 2.72\pm 0.46 & 2.62\pm 0.35\\
B/B_0=B(2)/B(3)        & 2.02\pm 0.56& 1.89\pm 0.23& 2.03 \pm 0.32& 2.02\pm 0.33 & 1.89\pm 0.22\\
F/F_0=F(2)/F(3)        & 1.16\pm 0.02& 1.18\pm 0.02& 1.16\pm 0.02& 1.16\pm 0.02  & 1.18\pm 0.02\\[0.8ex]
\hline
\hline\phantom{xx}&&&&&\\[-2.4ex]
\chi^2/N   & 28.7/11& 18.7/11 & 29.2/12& 29.4/12 & 19.2/12\\
{\rm Gaussian\ equiv.}  & 3.0 \sigma & 1.8\sigma  & 2.9\sigma  & 2.9\sigma  & 1.7\sigma 
\\[0.15ex]
\hline
\end{array}
$$
\caption{\small\it Results of fits performed on the data from 
RBC/UKQCD
collaboration on pseudoscalar masses and decay constants, including or not the topological susceptibility~\cite{Aoki:2010dy}. In two cases, finite-volume effects are taken into account.  In all cases, we considered only data with light pions 
and only statistical errors are shown.
The LECs are given at the scale $\mu =M_\rho$.
}\label{table:results0}
\end{table}

Focusing on the results including $\chi^{\rm pole}$ with finite-volume effects, we obtain the following convergence at $\chi^2_{\rm min}$ with the relative LO, NLO and HO contributions:
\begin{equation}
\begin{array}{rclcrcl}
F_\pi^2 & :& 0.65 + 0.40 -0.05\,,  & & F_\pi^2M_\pi^2 &:& 0.35+0.79-0.14\,,\\
F_K^2 &:& 0.47 + 0.56 -0.03\,,  & & F_K^2M_K^2 &:& 0.24+0.86 -0.10\,,\\
F_\eta^2 &:& 0.40 + 0.57 + 0.03\,,  & & F_\eta^2M_\eta^2 &:& 0.23+0.80 -0.03\,,\\
\chi^{\rm pole} &:& 0.22 + 0.63 + 0.15 \,,
\end{array}
\end{equation}
showing that the global convergence is good (small HO remainders), but the LO and NLO contributions are indeed competing numerically, confirming the results obtained in ref.~\cite{Bernard:2010ex}.

However, fitting the RBC/UKQCD data in the two volumes, we obtain a rather poor fit of
$\chi^2/$d.o.f.=29.4/12 without finite-volume
effects, which can be seen as a 2.9$\sigma$ discrepancy in a naive (purely Gaussian) statistical interpretation. The fit improves when one includes finite-volume effects, getting down to
a 1.7$\sigma$ effect, as can be seen by comparing the fits $A1$ and $A2$ (or $A4$ and $A5$) in Tab.~\ref{table:results0}. Indeed, even though these effects are rather small in the ranges of quark masses considered here, they tend to bring the various quantities in better agreement
with the lattice data (more on this issue can be found in App.~\ref{app:XiP}). The main contribution to $\chi^2_{\rm min}$ comes from $F_\pi$ (we will come back to this issue in the next section) whereas the topological susceptibility contributes only marginally.
 
In Fig.~\ref{figure:results0}, we illustrate the results of the fit for the topological susceptibility as a function of the simulated light-quark mass. Finite-volume effects have little impact, since at large $\tilde{m}$, the product of $\tilde{M}_\pi L$ is much larger than 1 and thus finite-volume effects are small, whereas at small $\tilde{m}$, the topological susceptibility (with or without finite volume effects) goes to 0 linearly. One notices also the linearity of the three curves, related to the fact that $q$ remains small ($\leq 0.4$) in this range of $\tilde{m}$, so that $\chi$ in eq~(\ref{eq:chinopole}) or (\ref{eq:chipole1})-(\ref{eq:chipole2}) can be expanded in powers of $q$ with only small $O(q^2)$ corrections. As is clear from eq.~(\ref{eq:chinopole}), the slopes of these curves at zero are not directly related to the three-flavour quark condensate, as it involves also the NLO LEC $L_6^r$ as well as chiral logarithms (once again, this feature is more related to the two-flavour quark condensate).

The authors of ref.~\cite{Aoki:2010dy} noticed that their value for the topological susceptibility obtained from the simulation with the lightest quark mass may suffer from \emph{``long auto-correlations in $Q$ that are not well resolved in [their] finite Markov chain of configurations''}. Indeed, the uncertainty attached to this particular point is much smaller than the other ones, which might indicate underestimated systematic effects. In addition, one notices that it is much lower than the other points, so that it would be very difficult to draw a smooth curve going through all the points as well as the origin (as required by the vanishing of the topological susceptibility when one quark mass is set to zero). 
 For these reasons, we did not include this data point in our analyses. We have however checked that including this additional point leads only to a worse $\chi^2$ without affecting the outcome for the LO parameters of the chiral Lagrangian in a significant way.

\begin{figure}[t!]
\begin{center}
\includegraphics[width=11cm,angle=0]{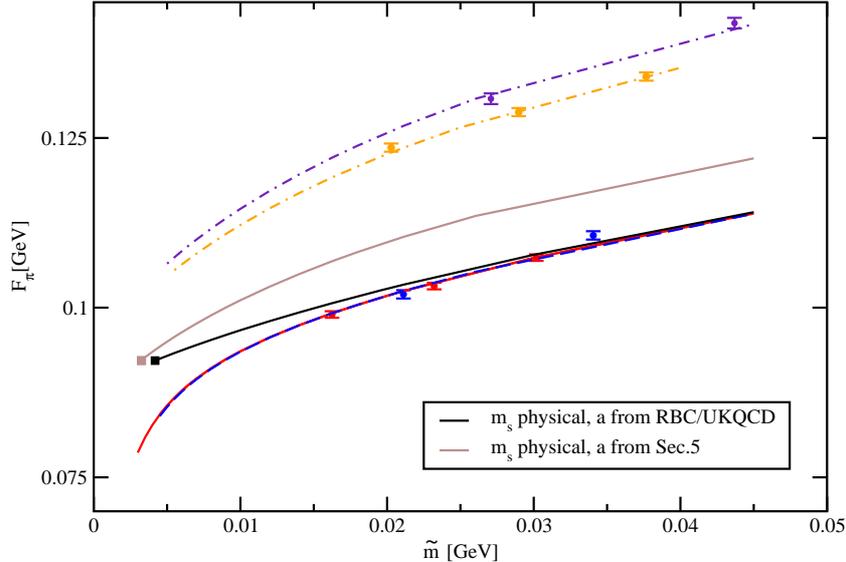}
\end{center}
\caption{\small\it The pion decay constant as a function of the light-quark mass. Blue [red] points correspond to the data for the
$(24)^3\times 64$ volume [$(32)^3\times 64$], using the estimates of lattice spacing given in ref.~\cite{Aoki:2010dy}. Purple [orange] points correspond to the same points, using our own estimates of lattice spacing discussed in Sec.~\ref{sec:fittodata}. The dashed curves indicate our best fit in each case ($A5$ and $B5$), as given in the last columns of tables 1 and 2 respectively. They are interrupted for $\tilde{M}_\pi L\leq 2$ where our description of  finite-volume effects becomes unreliable. The solid curves indicate the corresponding variations for the physical value of the strange quark mass, without finite-volume effects. In both cases (estimate of the lattice spacings from ref.~\cite{Aoki:2010dy} or Sec.~\ref{sec:latticespacing}), the position of the physical point $(m_{\rm phys}, F_\pi)$ is indicated with a square.
\label{figure:results1}}
\end{figure}

\section{The role of lattice spacing} \label{sec:latticespacing}

Our approach, allowing for a numerical competition between LO and NLO contributions to three-flavour chiral series, has provided a good, but not completely satisfying, fit of masses, decay constants and topological susceptibility from the RBC/UKQCD data.
This is illustrated in the lower part of Fig.~\ref{figure:results1}, where we plot $\tilde{F}_\pi$ as a function of $\tilde{m}$ (as given by the equivalent of eqs.~(\ref{etamasslatt})-(\ref{etadecaylatt}) for $\tilde{F}_\pi^2$ and $\tilde{F}_\pi^2\tilde{M}_\pi^2$, see Sec.~5.1 in ref.~\cite{Bernard:2010ex}). 
The (dashed) curves correspond to our best fit (last column in Tab.~\ref{table:results0}) and the solid line indicates the dependence of $F_\pi$ for a physical $m_s$ in an infinite volume. 
Indeed, in spite of this broad agreement, we notice that we get a large contribution to $\chi^2_{\rm min}$ from the $\tilde{F}_\pi$ data points. This is a reminder of the problem encountered in ref.~\cite{Aoki:2010dy}, where neither a chiral nor an analytic extrapolation formula was able
to accommodate the observed dependence of $\tilde{F}_\pi$ on $\tilde{m}$ with the physical point $(m_{\rm phys}, F_\pi)$. Our formalism can include both pieces of information with a reasonable 
$\chi^2_{\rm min}$, but we can improve the latter by letting
the physical value of $F_\pi$ vary as a free parameter. We obtain the results
indicated as fit $B1$ in the first column of Tab.~\ref{table:results1} with a very low value of $F_\pi=86.4$ MeV, in agreement with the results in ref.~\cite{Aoki:2010dy}.

This discrepancy between lattice and physical values of $F_\pi$ hints at a more general issue concerning the determination of the absolute scale 
of lattice quantities, which is obtained by converting the lattice spacing into physical units.
In ref.~\cite{Aoki:2010dy}, it was determined by identifying ``scaling trajectories'' corresponding to lines in the $(a,\tilde{m},\tilde{m}_s)$ space with fixed $\tilde{M}_\pi/\tilde{M}_\Omega$ and $\tilde{M}_K/\tilde{M}_\Omega$ ratios. An iterative interpolation method was used to reach values of $M_\pi, M_K, M_\Omega$ corresponding to physical values of their ratios, which was identified as the physical point for the quark masses. The lattice spacings were then determined by
requiring that $1/a=1.672/(a\tilde{M}_\Omega)$ GeV where 1.672 GeV is the
physical mass of this baryon and $a\tilde{M}_\Omega$ is the mass of the $\Omega$ as 
measured in lattice units. 
The actual interpolation was performed through
two kinds of interpolating formulae for the hadron masses in terms of quark masses, either based on 
NLO three-flavour $\chi$PT or an analytic (polynomial) ansatz, fitted to 
partially-quenched data (where sea- and valence-quark masses are different). This led to values of the lattice spacing that were compatible and quoted with an accuracy at the level of a few percents.

However, 
such determinations based on the quark-mass dependence of light-meson quantities might be affected significantly
if one takes into account a numerical competition between LO and NLO contributions to three-flavour chiral series. In this particular case, one should consider at the same time the chiral expansions of $F_P^2 M_P^2$ and $F_P^2$ and include HO remainders, to determine the dependence of $M_\pi$ and $M_K$ on the quark masses. As far as the $\Omega$ mass is concerned, we will follow ref.~\cite{Aoki:2010dy} and take a linear dependence on the quark masses~\footnote{The $\Omega$ mass was analysed as a function of $m_{u,d}$ in $N_f=2$ chiral perturbation theory in ref.~\cite{Tiburzi:2008bk}. The $\Omega^-$ field is an isoscalar under $SU(2)$, which prevents it from interacting with other baryons (contrary to other hyperons like $\Xi,\Sigma,\Lambda$) in two-flavour $\chi$PT. This makes the dependence of $M_\Omega$ on $M_\pi$ much simpler than for other nucleons, with a constant term supplemented with quadratic and quartic terms in $M_\pi$ (the quartic term including a chiral logarithm).}:
\begin{equation}\label{eq:momega}
\tilde{M}_\Omega=M_{\Omega}+c_1(\tilde{m}_s-m_{s})+c_2(\tilde{m}-m)\,,
\end{equation}
inspired by the analysis of RBC/UKQCD.
A fully consistent treatment should include a treatment of the baryon masses in our Re$\chi$PT framework, extending eq.~(\ref{eq:momega}) to chiral logarithms and HO remainders. Such an analysis, beyond the scope of the present article, is under way~\cite{wip} (in the present case, we have checked that adding a quartic term in eq.~(\ref{eq:momega}) does not change the results discussed below).

In addition, since we are interested in effects related to lattice spacing, we should also consider discretisation errors, which could reach 10-15\% in the analysis of ref.~\cite{Aoki:2010dy}. We follow the latter analysis and include only leading-order discretisation effects affecting the decay constants~\footnote{In principle, one should consider all the terms coming from discretisation effects and due to the breaking of chiral symmetry, and also add correction terms for the masses. This would however increase the number of free parameters and lead to fits with a poorly stability, due to flat directions in the subspace of correction terms and the limited number of data points.}:
\begin{eqnarray}
\tilde{F}_\pi^2(a,V,\tilde{m},\tilde{m_s})&=&\tilde{F}_\pi^2(0,V,\tilde{m},\tilde{m_s})+F_\pi^2 Z(3) c_{F_\pi} a^2\,,\\
\tilde{F}_K^2(a,V,\tilde{m},\tilde{m_s})&=&\tilde{F}_K^2(0,V,\tilde{m},\tilde{m_s})+F_\pi^2 Z(3) c_{F_K} a^2\,,
\end{eqnarray}
where the correction term is defined with respect to the leading-order term in the chiral expansion.

We are not in a position to perform the same joint determination of lattice spacings and quark masses as the RBC/UKQCD collaboration to include the dependence of the pion and kaon masses on the light-quark mass inferred from Re$\chi$PT from scratch. However,
we can perform a combined fit of the pion and kaon masses and decay constants, as well as the $\Omega$ mass (collected in App.~\ref{app:data}), with the following parameters:
\begin{itemize}
\item the three leading-order parameters $r,X(3),Z(3)$,
\item a reference ratio between a simulated strange quark (chosen conventionally to be for the $24^3$ simulations) and the physical strange quark mass,
\item the HO remainders associated with the pion and kaon masses and decay constants (denoted $d,e,d',e'$),
\item the ratio $F_K/F_\pi$ (on the other hand, we set $F_\pi=92.2$ MeV),
\item if the lattice spacings are left free, the two effective constants $c_1,c_2$ for the $\Omega$ mass,
\item if discretisation errors are included, the two effective constants $c_{F_\pi},c_{F_K}$ for the decay constants,
\item if the topological susceptibility is included, the corresponding HO remainders for the $\eta$ mass and decay constant ($d_\eta,e_\eta$) as well as the one for the topological susceptibility ($d_\chi$).
\end{itemize}
Our approach is not very different in its spirit from the ``combined scaling and chiral fitting'' performed in ref.~\cite{Aoki:2010dy}, up to the following modifications: we include information on the values of the masses and decay constants at the physical point, we consider $F_P^2M_P^2$ and $F_P^2$ rather than $M_P^2$ and $F_P$, we use three-flavour Re$\chi$PT rather than two-flavour expansions to perform the interpolation of the data, we include the presence of HO remainders, we do not include partially-quenched data, we fix at the same time $m,m_s$ and the lattice spacings in contrast with the two-step procedure in ref.~\cite{Aoki:2010dy} (physical masses first, lattice spacing afterwards).

\begin{table}[ht!]
$$
\begin{array}{|c||c||c|c|c||c|}
\hline
{\rm Fit\ name} & B1 & B2 & B3 & B4 & B5\\
\hline
\chi & {\rm No}  & {\rm No} & {\rm No} & {\rm No} &  \chi^{\rm pole}\\
{\rm Finite\ volume} &  {\rm Yes} & {\rm Yes} &  {\rm Yes} & {\rm Yes} & {\rm Yes} \\
{\rm Lattice\ spacing} &     {\rm Fixed}               & {\rm From\ }M_\Omega & {\rm Fixed}    
& {\rm From\ }M_\Omega& {\rm From\ }M_\Omega \\
O(a^2){\rm \ corr} \ &      {\rm No}             & {\rm No} & {\rm Yes}   & {\rm Yes}& {\rm Yes}  \\
\hline
F_\pi [{\rm MeV}] & 86.5\pm 2.2 & 92.2^\star & 92.2^\star & 92.2^\star & 92.2^\star \\
\hline\hline
\phantom{xx}&&\\[-2.9ex]
r       & 25.6 \pm 1.1  & 26.1\pm 1.1   & 23.3\pm0.8 & 25.6\pm 1.1 &  25.6\pm 1.0 \\
X(3)    & 0.40 \pm 0.04 & 0.54\pm 0.05  & 0.35 \pm 0.07 & 0.55\pm 0.04 & 0.55\pm 0.04\\
Y(3)    & 0.79 \pm 0.10 & 0.86 \pm 0.08   & 0.52 \pm 0.05 & 0.91\pm 0.06 & 0.91\pm 0.07\\
Z(3)    & 0.50 \pm 0.07 & 0.62 \pm 0.05   & 0.68 \pm 0.07 & 0.60\pm 0.05 & 0.60\pm 0.04 \\
F_K/F_\pi & 1.23\pm 0.03& 1.25 \pm 0.03   & 1.17 \pm 0.02 & 1.25 \pm 0.02 & 1.25\pm 0.02 \\
{\rm Rem.\ at\ limit} & - &  d & e & d,e & d,e,d_\eta,e_\eta,d_\chi\\
\tilde{m}_{s}(24^3)/m_s & 1.14\pm 0.03 & 1.62 \pm 0.16   & 1.12\pm 0.03 & 1.71\pm 0.13 & 1.71\pm 0.13\\
a^{-1}(24^3) [{\rm GeV}] & 1.73^\star& 2.14 \pm 0.12   & 1.73^\star & 2.02 \pm 0.08 & 2.19\pm 0.08\\
a^{-1}(32^3)  [{\rm GeV}] & 2.28^\star& 2.76 \pm 0.13   & 2.28^\star & 2.82 \pm 0.09 & 2.81\pm 0.09 \\
c_1 & - & 9.08 \pm 0.44  & - & 8.98 \pm 0.37 &  8.98\pm 0.37 \\
c_2 & - & 1.42 \pm 0.43 & - & 1.44\pm 0.44 & 1.44\pm 0.43 \\
c_{F_\pi}\ [{\rm GeV}^2]  & - & - &  0.13 \pm 0.14 & -1.09 \pm 1.19 & -1.09\pm 1.19\\
c_{F_K}\ [{\rm GeV}^2]& - & - & 0.13 \pm 0.17  & 0.09 \pm 0.48 & 0.09\pm 0.48 \\
\hline
\phantom{xx}&&\\[-2.9ex]
m_s(2\ {\rm GeV}) [{\rm MeV}] & 96.5\pm 2.2 & 83.9\pm 8.1   & 97.9 \pm 2.2 & 81.2 \pm 6.0 & 81.2\pm 6.0 \\
m(2\ {\rm GeV})   [{\rm MeV}] & 3.8\pm 0.2  & 3.2 \pm 0.4   & 4.2 \pm 0.1 & 3.2\pm 0.3 & 3.2\pm 0.3  \\
\Sigma_0^{1/3}(2\ {\rm MeV}) [{\rm GeV}] & 197\pm 7&  240\pm 16    & 191\pm 2 & 243\pm 12 & 243\pm 12 \\
B_0(2\ {\rm GeV}) [{\rm GeV}] & 2.03\pm 0.32& 2.60\pm 0.52   & 1.21 \pm 0.12 & 2.81\pm 0.39 & 2.81\pm 0.38\\
F_0 [{\rm MeV}]               & 61.5\pm 5.2 & 72.8\pm 2.8   & 76.0 \pm 3.7 & 71.3 \pm 2.4 & 71.3\pm 2.4\\
F_\eta [{\rm MeV}]              & -  &  -   & - & - &  123\pm 3  \\
\chi^{\rm pole}\cdot 10^4 [{\rm GeV}^4]  & - & 0.33\pm 0.03  & 0.41 \pm 0.03 & 0.33\pm 0.03 & 0.43 \pm 0.03 \\ 
\tilde{m}_{s}(32^3)/m_s & 1.09 \pm 0.03 & 1.52\pm 0.15   & 1.08 \pm 0.03 & 1.61 \pm 0.12 & 1.61\pm 0.12 \\
\hline
\phantom{xx}&&\\[-2.9ex]
 L_4(\mu)\cdot 10^3   & 0.28 \pm 0.29 & -0.11 \pm 0.14   & 0.99 \pm 0.17 & -0.08 \pm 0.09 & -0.08\pm 0.09 \\
 L_5(\mu)\cdot 10^3   & 1.62 \pm 0.19 & 2.13\pm 0.21  & 2.03 \pm 0.43 & 2.00 \pm 0.20 & 2.01\pm 0.20 \\
 L_6(\mu)\cdot 10^3   & 0.47 \pm 0.35 & 0.09\pm 0.10   & 2.53 \pm 0.79 & 0.05 \pm 0.07 & 0.04\pm 0.07 \\
 L_7(\mu)\cdot 10^3   & - & -  & - & - & -0.25\pm 0.14\\
 L_8(\mu)\cdot 10^3   & 1.08 \pm 0.38 & 1.13\pm 0.20   &  3.42 \pm 0.88 & 1.04\pm 0.14 & 1.04\pm 0.09 \\[0.15ex]
 \hline
 \phantom{xxx}&&\\[-2.9ex]
X(2)  & 0.89 \pm 0.02 & 0.89\pm 0.01  & 0.90 \pm 0.01 & 0.88 \pm 0.01 & 0.88\pm0.01 \\
Y(2)  & 1.03 \pm 0.02 & 1.03\pm 0.01 & 0.99 \pm 0.01 & 1.04 \pm 0.01 &1.04\pm 0.01 \\
Z(2)  & 0.86 \pm 0.02 & 0.86 \pm 0.01   & 0.91 \pm 0.01 &  0.86\pm 0.01 & 0.85\pm 0.01 \\
\Sigma^{1/3}(2\ {\rm GeV}) [{\rm MeV}] & 258\pm 4 & 248\pm 2   & 261 \pm 3 & 285 \pm 7 & 285 \pm 8\\
B (2\ {\rm GeV}) [{\rm GeV}] & 2.68\pm 0.16 &3.13 \pm 0.42   & 1.90 \pm 0.18 & 3.18 \pm 0.28 & 3.18\pm 0.27 \\
F [{\rm MeV}]       & 80.2\pm 2.9 & 85.5\pm 0.50   & 88.0 \pm 0.4& 85.2 \pm 0.4 & 85.2\pm 0.4\\
\bar\ell_3     & 4.0\pm 1.5& 4.5\pm 0.9   & -1.0 \pm 0.9 & 4.6 \pm 0.7 & 4.6\pm 0.7\\
\bar\ell_4     & 4.2\pm 0.4& 4.7\pm 0.3   & 3.1 \pm 0.3 & 4.9 \pm 0.3 & 4.9\pm 0.3 \\
 \hline
 \phantom{xxx}&&\\[-2.9ex]
\Sigma/\Sigma_0=\Sigma(2)/\Sigma(3)& 2.24\pm 0.22 & 1.66\pm 0.14   & 2.54 \pm 0.02 & 1.61 \pm 0.11 & 1.61 \pm 0.11\\
B/B_0 =B(2)/B(3)         & 1.32\pm 0.17 & 1.20 \pm 0.10   & 1.90 \pm 0.18 & 1.13 \pm 0.09 &1.13\pm 0.08 \\
F/F_0 =F(2)/F(3)         & 1.30\pm 0.08 & 1.17\pm 0.05   & 1.15 \pm 0.06 &  1.19\pm 0.04 & 1.19 \pm 0.04\\[0.8ex]
\hline
\hline\phantom{xx}&&&\\[-2.9ex]
\chi^2/N & 12.1/10 & 17.3/12 & 18.1/9 & 16.0/10 &  16.2/11\\
{\rm Gaussian\ equiv.} & 1.1 \sigma & 1.5\sigma  & 2.1 \sigma & 1.6\sigma & 1.5\sigma  \\[0.15ex]
\hline
\end{array}
$$
\caption{\small\it Results of fits performed on the data from 
RBC/UKQCD
collaboration on pseudoscalar masses, decay constants and topological susceptibility~\cite{Aoki:2010dy}. The first fit lets $F_\pi$ vary freely. The other columns either a determination of the lattice spacings using the $\Omega$ mass or $O(a^2)$ discretisation effects for the decay constants. The star superscript indicates a quantity set to a fixed value (no uncertainty). These results should be compared with fits $A2$ and $A5$ in table 1.
}\label{table:results1}
\end{table}

\clearpage

The results are given in Tab.~\ref{table:results1} including
finite-volume effects~\footnote{Contrary to Sec.~\ref{sec:fittodata}, including or not finite-volume effects affects the value of  $\chi^2_{\rm min}$ only marginally, because the small changes due to finite-volume effects can be mimicked at the level of the fit by a slight shift in the values of the lattice spacings. This was also observed in the fits of the RBC/UKQCD collaboration on their own results~\cite{Aoki:2010dy}.}. 
First we consider fits $B2,3,4$ without the topological susceptibility, including or not discretisation errors and lattice spacing determined from $M_\Omega$. Discretisation errors remain generally small (at most 5 \%) apart for $F_\pi^2$ in fit $B4$ (20\% effect, in good agreement with the results of ref.~\cite{Aoki:2010dy}), but lead to enlarged uncertainties on the other parameters. We notice that these discretisation effects are compatible with zero within error bars, which explains that the fits $A2$ and $B3$ (differing only through the effect of discretisation errors) yields very similar results. The (re)determination of lattice spacing through $M_\Omega$ performed in $B2$ and $B4$ has a much more significant impact, as it tends to decrease the lattice spacings significantly (20-30\%) as well as the value of the physical quark masses, and to increase $F_\pi/F_K$ noticeably. In these fits, the simulated quark masses stand much further away from the physical value than quoted in ref.~\cite{Aoki:2010dy}, implying that HO remainders at the simulated quark masses (scaling generically as $p^2$ with respect to the HO at the physical point) play a significant role in the chiral expansion of observables (up to 40\% for the heaviest $\tilde{m}_s$). The large error bars for dimensionful quantities is a reflection of the uncertainty on the lattice spacings determined from $M_\Omega$.
The significant difference between the value obtained the lattice spacings obtained here and in ref.~\cite{Aoki:2010dy} 
suggests a reanalysis the lattice data following the same procedure as the RBC/UKQCD collaboration (with the determination of the scaling trajectories and an iterative determination of masses and lattice spacings) but using the Re$\chi$PT formulae instead of analytic or standard $\chi$PT ones. Such a study would be very welcome to establish the effect observed in our fits, but is clearly beyond the scope of the present article.

Finally, we include the topological susceptibility among the inputs in fit $B5$. As in Sec.~\ref{sec:fittodata},
the role of this input in the fit is marginal compared to the other data, and the outcome of fits $B4$ and $B5$ is very similar. Indeed, the $\chi^2_{\rm min}$ obtained from fit $B5$ gets similar contributions from $F_\pi$, $F_\pi M_\pi$, $F_K M_K$ and $M_\Omega$, whereas the contribution from the topological susceptibility is small. The rest of the analysis is unchanged, with a competition between LO and NLO contributions for the observables of interest:
\begin{equation}
\begin{array}{rclcrcl}
 F_\pi^2&:&  0.60 + 0.25 + 0.15\,, &&  F_\pi^2M_\pi^2&:&  0.55 + 0.30 + 0.15\,,\\
 F_K^2&:&  0.38 +  0.52 + 0.10\,, &&  F_K^2M_K^2&:&  0.37 + 0.52 + 0.11\,,\\ 
 F_\eta^2&:&  0.33 + 0.58 + 0.09\,, &&  F_\eta^2M_\eta^2&:& 0.35 + 0.66 - 0.01\,,\\
 \chi^{\rm pole} &:& 0.50 + 0.35 + 0.15 \,.
\end{array}
\end{equation}
The competition between LO and NLO terms  of the chiral expansions observed in the previous fits  remains though a bit less pronounced:  $Y(3)$ is closer to one, and the {\rm value of $L_6$ is closer to zero, but there is still} an enhancement of NLO contributions to observables. Indeed, this enhancement  is parametrised by $m_s B_0\Delta L_4$ and $m_s B_0 \Delta L_6$ rather than $L_4$ and $L_6$ themselves, as can be seen
for instance in eqs.~(\ref{etadecay})-(\ref{etamass})
(with the typical values $\Delta L_4=L_4^r(M_\rho)+0.50\cdot 10^{-3}$ and   $\Delta L_6=L_6^r(M_\rho)+0.25\cdot 10^{-3}$ for $r=25$ and $Y(3)=1$)~\cite{DescotesGenon:2000di,DescotesGenon:2000ct}. 

Our fit $B5$ features a fairly good $\chi^2_{\rm min}/N$ compared to the results in ref.~\cite{Aoki:2010dy}, taking into account that our fit reproduces the physical values of the pseudoscalar masses and $F_\pi$ by construction. We agree with ref.~\cite{Aoki:2010dy} on the size of discretisation errors, but obtain different results concerning the determination of the lattice spacings from $M_\Omega$. The main difference stems from our use of Re$\chi$PT in the fit, as can be illustrated by performing the same fit as $B_5$ but constraining $X(3)$ and $Z(3)$ to remain between 0.9 and 1.05. This last constraint mimics the usual assumption made in three-flavour $\chi$PT that both $F_\pi$ and $M_\pi$ are nearly saturated by their LO term. The minimum of the fit may look satisfying with $r=26.1$, $X(3)=0.9$, $Y(3)=0.97$, $Z(3)=0.93$, $F_K/F_\pi=1.18$ and values of the lattice spacings compatible at the 5\% level with ref.~\cite{Aoki:2010dy}, but the value at the minimum is rather awful with $\chi^2_{\rm min}=151/11$. This shows clearly that allowing for a numerical competition between LO and NLO is mandatory to reach decent fits to the whole set of data considered here.

\begin{figure}[t!]
\begin{center}
\includegraphics[width=11cm,angle=0]{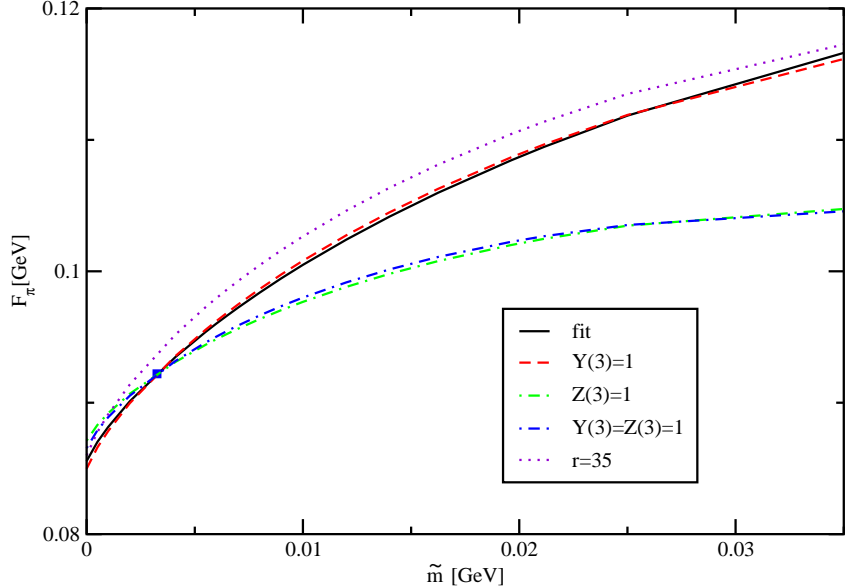}
\end{center}
\caption{\small\it The pion decay constant as a function of the light-quark mass, according to the best fit in the last column of Tab.~\ref{table:results1} (solid), with the same inputs but setting $Y(3)=1$ and/or $Z(3)=1$ (dashed and dashed-dotted curves), and with the same inputs but setting $r=35$ (dotted curve).
The physical point according to the fit result is indicated with a square.
\label{figure:curve}}
\end{figure}

The results from fit $B5$ are also illustrated in the upper part of Fig.~\ref{figure:results1}.
The change in the value of the lattice spacings rescales all lattice data points, which
fall on the dashed curves corresponding to the best Re$\chi$PT fit including finite-volume effects (responsible for most of the curvature at small $\tilde{m}$) more easily than in Sec.~\ref{sec:fittodata}. 
The values of $p=\tilde{m}_s/m_s$ and $Y(3)$ are larger than in fit $A5$: from eq.~(\ref{etadecay}) (which has a similar structure to the Re$\chi$PT expression for  $\tilde{F}_\pi$), we see that NLO contributions
are proportional to the product $p Y(3)$ and tend thus to drive the (dashed) curves corresponding to the two volumes further apart than in the previous section.
We also display the (solid) curve corresponding to the physical value of $m_s$ in an infinite volume, which by construction passes through the physical point $F_\pi=92.2$ MeV (this point corresponds to a slightly different value of $\tilde{m}$ from Sec.~\ref{sec:fittodata}). 
The curvature is then essentially tied to the value at the origin, i.e., $Z(2)$: a lower value of $Z(2)$ will yield a steeper increase of $\tilde{F}_\pi$ when $\tilde{m}$ increases. The dependence of $Z(2)$ on the various parameters can be read from the three-flavour expansion of $F_\pi^2$ in the chiral limit $m_u=m_d=0$~\cite{DescotesGenon:2003cg}:
\begin{equation}
Z(2)=\frac{r}{r+2}[1-\eta( r)]+\frac{2}{r+2}Z(3)-\frac{r}{r+2}Y(3)g_1+\ldots
\end{equation}
where $g_1$ denotes a small positive combination of chiral logarithms (around 7\% near the physical point) and the ellipsis indicate HO remainders.
Therefore, a dependence with a stronger curvature around the physical point and a flatter behavior above can be achieved
by taking a larger value $Y(3)$ or a smaller value of $Z(3)$, as illustrated in Fig.~\ref{figure:curve} and observed in fit $B5$ compared to $A5$. As an illustration, we indicate on the same figure the curves obtained with the same inputs as the best values for fit $B5$ of table~\ref{table:results1}, but setting $Y(3)=1$ and/or $Z(3)=1$. The case of a complete saturation of the chiral series for $F_\pi^2$ and $F_\pi^2M_\pi^2$ by their LO contribution ($X(3)=Z(3)=1$) yields a higher $Z(2)$ and a flatter curve than our best fit. 
Finally, we notice that an increase of $r$ (at fixed $p$) yields a slight increase of $Z(2)$, but 
more importantly an upward shift of the physical value of $\tilde{m}$, so that the corresponding line remains above the best-fit curve over a larger range of $\tilde{m}$.
The combination of these effects allows our formulae to reproduce a diversity of behaviours for the dependence of $\tilde{F}_\pi$ on $\tilde{m}$, including the one exhibited by the RBC/UKQCD data.

Since the results presented in table~\ref{table:results0} do not include effects related to lattice spacings and since the determination of the lattice spacings involves assumptions on the form of chiral extrapolation, we consider the most complete fit $B5$ in table~\ref{table:results1} as the final result of our analysis, featuring a satisfying $\chi^2_{\rm min}/N$.
Interestingly, these results show a good compatibility with what was obtained in ref.~\cite{Bernard:2010ex} for the parameters of the LO chiral Lagrangian according to the results of the PACS-CS collaboration~\cite{Aoki:2008sm}.
The value of the  condensate in the $N_f=3$ chiral limit is 
\begin{equation}
(\Sigma(3;2\,{\rm GeV}))^{1/3} = 243 \pm 12 \, {\rm MeV}\,,
\end{equation}   
on the lower side of was obtained by the RBC/UKQCD collaboration (i.e., $256\pm 6$ MeV)~\cite{Aoki:2010dy}, while the condensate in the two-flavour limit is
\begin{equation}
(\Sigma(2;2\,{\rm GeV}))^{1/3} = 285\pm 8 \,{\rm MeV}\,, \qquad
\Sigma(2;2{\rm GeV})/\Sigma(3;2{\rm GeV})=1.51\pm 0.11\,,
\end{equation}
which illustrates the paramagnetic suppression of the $N_f=3$ condensate with respect to the $N_f=2$ one (a similar statement holds for the pseudoscalar decay constant).

\section{Conclusion}
We have studied the topological susceptibility, a very interesting quantity related to chiral properties of QCD vacuum, and in particular to the quark condensate. This has led lattice collaborations to use this observable to determine the quark condensate, in addition to studies of the spectrum and dynamics of the light pseudoscalar mesons. However, it is important to assess higher-order corrections to the deceivingly simple connection between the topological susceptibility and the three-flavour quark condensate at leading order.
Following our recent work on potential issues in three-flavour chiral extrapolations of lattice data~\cite{Bernard:2010ex}, we have reassessed the information that can be extracted from this quantity allowing for a significant paramagnetic suppression of the $N_f=3$ quark condensate using the Resummed $\chi$PT (Re$\chi$PT) framework. We have noticed that for lattice simulations close to the physical situation (with a significant mass hierarchy between the dynamical strange and $u,d$ quarks), the topological susceptibility would involve essentially the same combination of low-energy constants as the two-flavour quark condensate, and thus would not provide access to the three-flavour quark condensate as naively expected. In particular, using the leading-order three-flavour $\chi$PT formula would be particularly misleading if there is indeed a significant paramagnetic suppression of the three-flavour quark condensate.

In order to escape this problem, two alternatives can be considered. A first possibility consists in  performing further lattice simulations with hierarchy of quark masses different from the physical case, for instance with three dynamical quark masses of similar masses, or with a very significant strong isospin breaking. A second possibility relies on the combination of the topological susceptibility together with other sources of information on three-flavour chiral symmetry breaking, such as the spectrum of pseudoscalar mesons. Following this path, we have focused on
RBC/UKQCD recent results~\cite{Aoki:2010dy} which provide data on pseudoscalar masses and decay constants as well as the topological susceptibility for two different volumes.
Analysing the RBC/UKQCD unitary data points and including finite-volume effects, we have 
confirmed that these data do suggest such a suppression of the leading order $N_f=3$ chiral order parameters (quark condensate and pseudoscalar decay constant) and the enhancement of next-to-leading order contributions related to the violation of the Zweig rule in the scalar sector ($L_4$ and $L_6$ contributions). The outcome of this first series of fits is mainly driven by the input from pseudoscalar masses and decay constants, and the data on topological susceptibility shows a good compatibility with the pattern of three-flavour chiral symmetry breaking already found in ref.~\cite{Bernard:2010ex}. In addition, we confirm the difficulties (though at a lesser degree) encountered by the RBC/UKQCD collaboration to accommodate the dependence of $F_\pi$ on the light-quark mass given by their data and the physical value of $F_\pi$ at the same time. 

This problem has led us to
reconsider the procedure used to determine the lattice spacing using our expressions to describe the dependence of the pseudoscalar observables on the quark masses. We were not able to
follow the same procedure as RBC/UKQCD for the determination of the lattice spacings (based on the determination of scaling trajectories and the dependence of the $\Omega$ baryon on quark masses). However, we performed a joint fit of 
pion, kaon and $\Omega$ observables to fix the lattice spacings. We also considered leading-order discretisation errors that may affect the kaon and pion decay constants. We finally performed a fit combining these two effects and adding data on the topological susceptibility. We obtained thus our final results given by the fit $B5$ in table~\ref{table:results1}. We noticed
 a significant enhancement (20\%-30\%) of the inverse of lattice spacings compared to the values quoted in ref.~\cite{Aoki:2010dy}, as well as small discretisation errors (5\% or less, and compatible with zero).
Like in the previous analysis, the data on the topological susceptibility play a marginal role, but show a good compatibility with the rest of the data, yielding a satisfying $\chi^2_{\rm min}/N$. Since they include the largest sets of data and use Re$\chi$PT consistently for the fit of the data and the determination of the lattice spacings, the results of fit $B5$ should thus be considered as the actual outcome of our analysis.
  
The impact of our analysis on the determination of the lattice spacings is remarkable, and it calls for a confirmation by reanalysing the lattice data following the same procedure as the RBC/UKQCD collaboration (with the determination of the scaling trajectories and an iterative determination of masses and lattice spacings) but using the Re$\chi$PT formulae. However such a cross-check would go beyond the scope of the present article and our abilities.
Let us add hat the inclusion of these effects does not modify the emerging picture of three-flavour chiral symmetry breaking already obtained in ref.~\cite{Bernard:2010ex}, with a significant competition between leading and next-to-leading orders in the chiral series for masses, decay constants and topological susceptibility.

We have used the RBC/UKQCD data as a case study for chiral extrapolations on the lattice, including both the pseudoscalar spectrum and the topological susceptibility, which can (and should) be extended to the results of other lattice collaborations.
Our present analysis suggests also the inclusion of additional observables, e.g., other topological quantities such as the topological coefficient $c_4$ as well as baryon observables, in our framework~\cite{wip}. But even before considering this extension, the generation of  
further data points at lower quark masses would help clarifying the issue of the determination of lattice spacing as well as confirming the pattern of three-flavour chiral symmetry breaking emerging from our studies, with significant contributions from strange sea quarks leading to a non-trivial structure of $N_f=3$ chiral expansions.

\section*{Acknowledgments}

We would like to thank O.~P\`ene and S. Beane for useful discussions on several aspects of  lattice simulations.

\appendix

\section{NLO low-energy constants} \label{app:lec}

As recalled in ref.~\cite{Bernard:2010ex} and outlined in Sec.~\ref{sec:rechpt}, the exact mass and decay constant identities obtained within the Re$\chi$PT framework for $F_\pi^2$, $F_K^2$, $F_\pi^2M_\pi^2$ and $F_K^2M_K^2$ can be inverted to reexpress NLO LECs in terms of LO parameters of the chiral Lagrangian, physical quantities, and HO remainders:
\begin{eqnarray} 
\label{eq:deltal4}
Y(3)\Delta L_4 &=& \frac{1}{8(r+2)}\frac{F_\pi^2}{M_\pi^2}
  [1-\eta(r)-Z(3)-e]\,,
\\ 
\label{eq:deltal5}
Y(3)\Delta L_5 &=& \frac{1}{8}\frac{F_\pi^2}{M_\pi^2}
  [\eta(r)+e']\,, 
\\
\label{eq:deltal6}
Y^2(3)\Delta L_6 &=& \frac{1}{16(r+2)}\frac{F_\pi^2}{M_\pi^2}
  [1-\epsilon(r)-X(3)-d]\,,
\\ 
\label{eq:deltal8}
Y^2(3)\Delta L_8 &=& \frac{1}{16}\frac{F_\pi^2}{M_\pi^2}
  [\epsilon(r)+d']\,.
\end{eqnarray}
with
$\epsilon(r)$ defined in eq.~(\ref{funcr}), $d$, $d'$, $e$ and $e'$ combinations of  
remainders associated with the chiral expansions of $\pi$, $K$ masses and decay 
constants respectively and 
\begin{equation}  
\eta(r)=\frac{2}{r-1}\left(\frac{F_K^2}{F_\pi^2}-1\right)\,,
\end{equation}
$\Delta L_i=L_i^r(\mu)-\hat{L}_i(\mu)$ is independent of the renormalisation scale $\mu$
and combine the (renormalized and quark-mass independent)
constants $L_{4,5,6,8}$ together with chiral logarithms:
\begin{eqnarray}
32\pi^2\hat{L}_4(\mu)
          &=& \frac{1}{8}
          \log\frac{\mo_K^2}{\mu^2}
  -\frac{1}{8(r-1)(r+2)}
  \left[(4r+1)\log \frac{\mo_K^2}{\mo_\pi^2} 
      + (2r+1)\log \frac{\mo_\eta^2}{\mo_K^2}  \right]   
\label{eq:l4} \,,\\
32\pi^2 \hat{L}_5(\mu) &=& \frac{1}{8}
    \left[\log\frac{\mo_K^2}{\mu^2}+2\log\frac{\mo_\eta^2}{\mu^2}\right]
+\frac{1}{8(r-1)}
    \left[3\log\frac{\mo_\eta^2}{\mo_K^2}+5\log\frac{\mo_K^2}{\mo_\pi^2}\right]
\label{eq:l5} \,,\\
32\pi^2\hat{L}_6(\mu) &=& \frac{1}{16}\left[
       \log\frac{\mo_K^2}{\mu^2} 
       + \frac{2}{9}\log \frac{\mo_\eta^2}{\mu^2}
         \right] 
 -\frac{1}{16}\frac{r}{(r+2)(r-1)} \left[ 3 \log
          \frac{\mo_K^2}{\mo_\pi^2}  + \log \frac{\mo_\eta^2}{\mo_K^2} \right]
\label{eq:l6}\,, \\
32\pi^2 \hat{L}_8(\mu) &=& \frac{1}{16}
    \left[\log\frac{\mo_K^2}{\mu^2}+\frac{2}{3}\log\frac{\mo_\eta^2}{\mu^2}\right]
+\frac{1}{16(r-1)} \left[ 3 \log
          \frac{\mo_K^2}{\mo_\pi^2}  + \log \frac{\mo_\eta^2}{\mo_K^2} \right]
\label{eq:l8}\,.
\end{eqnarray}
where $\mo_P^2$ are the LO contributions to the pseudoscalar masses, see eq.~(\ref{eq:kring}).

\section{Lattice inputs} \label{app:data}

We take our data points for pseudoscalar decay constants and masses as well as from topological susceptibility 
from the recent work of the RBC/UKQCD collaboration~\cite{Aoki:2010dy}. They considered 2+1 dynamical flavours of domain wall fermions for two different lattice volumes $24^3\times 64\times 16$
and $32^3\times 64\times 16$ (where the 16 corresponds to the extent of the fifth dimension inherent in the domain-wall fermion formulation of QCD).
We consider only unitary sets where the masses of the sea and valence quarks are identical, with parameters recalled in Tab.~\ref{tab:data1} and observables
in Tab.~\ref{tab:data2}.

\begin{table}
\begin{center}
\begin{tabular}{cccccccccccc}
$L$ & $a^{-1}$ & $\Delta_q$ & $\Delta_s$ & $Z_m^q$ & $Z_m^s$ & 
  $\tilde{m}_q$ & $\tilde{m}_s$ & $m_s$ & $p$ & $q$ \\
  \hline
$24$ & 1.73 & 0.005 & 0.04 & 1.4980 & 1.4707 & 0.0211 & 0.1098 & 0.0962 &  1.1420 & 0.1924 & 
 \\
 & & 0.010 & & & &  0.0342 & &  & & 0.3111   \\ \hline
$32$ & 2.28 & 0.004 & 0.03 & 1.527 & 1.510 & 0.0163 & 0.1056 & 0.0962 & 1.0976 & 0.1539 \\
& & 0.006 & & & & 0.0232 & & & & 0.2196 \\
& & 0.008 & & & & 0.0302 & & & & 0.2859 
\end{tabular}
\caption{\small\it Parameters of the unitary lattice sets taken from ref.~\cite{Aoki:2010dy}.
\label{tab:data1}}
\end{center}
\end{table}

We denote $\Delta_q=a(\tilde{m}_q-m_{res})$ the combination corresponding to bare masses (before addition of the residual mass $m_{res}$ and the conversion into the MS-bar scheme by a multiplication by $Z^m$). We give $a^{-1}$ and quark masses in units of GeV, $\chi$ in units of $10^{-4}\ {\rm GeV}^4$, $F_P^2$ in units of $10^{-3}\ {\rm GeV}^2$,
$F_P^2 M_P^2$ in units of $10^{-3}\ {\rm GeV}^4$. The dimensionful quantities have been converted from the lattice results by multiplying by the appropriate power of the lattice spacing, assuming for the latter the values quoted in the table. When the lattice spacings are allowed to vary and included in the parameters of the fits in Sec.~\ref{sec:latticespacing}, these quantities are naturally rescaled by the appropriate power of the relevant lattice spacing. 

\begin{table}
\begin{center}
\begin{tabular}{ccccccccccccc}
$L$ & $p$ & $q$ & $\chi \cdot 10^{-4}$ & $F_\pi^2$ & $F_K^2$ & $F_\pi^2M_\pi^2$ & $F_K^2M_K^2$\\
\hline
$24$ &  1.1420 & 0.1924 & 
2.90(14)& 10.41(13) & 13.47(14) & 1.129(16) & 4.459(50)\\
 & & 0.3111 & 4.58(25) & 12.26(14)& 14.85(18) & 2.152(25) & 5.471(68)\\ \hline
$32$ & 1.0976 & 0.1539 & [1.48(14)]& 9.799(96)& 12.99(11)& 0.820(10) & 3.988(36)\\
& & 0.2196 &  3.22(55)& 10.65(10) & 13.51(11) & 1.265(13) & 4.390(38)\\
& & 0.2859 &  3.63(72)& 11.53(10) & 14.16(12) & 1.788(18) & 4.895(42)
\end{tabular}
\caption{\small\it Lattice data for the pseudoscalar masses and decay constants as well as topological susceptibility taken from ref.~\cite{Aoki:2010dy}.
\label{tab:data2}
}
\end{center}
\end{table}

In Tab.~\ref{tab:data2}, we do not include uncertainties coming from the determination of the lattice spacings, as these uncertainties would be completely correlated. 
Moreover, as explained at the end of Sec.~\ref{sec:fittodata}, we do not include the value of the topological susceptibility at a lighter quark mass given in ref.~\cite{Aoki:2010dy}, since it is likely to be affected by large systematics of unknown origin.
In Sec.~\ref{sec:latticespacing}, we discuss the determination of the lattice spacings performed in ref.~\cite{Aoki:2010dy} using the mass of the $\Omega$ baryon, gathered in Tab.~\ref{tab:data3}.

\begin{table}
\begin{center}
\begin{tabular}{cccccccccccc}
$L$ & $a^{-1}$ & $\Delta_q$ & $\Delta_s$ & $a M_\Omega$ \\
\hline
$24$ & 1.73 & 0.005 & 0.04 & 1.013(3)
 \\
 & & 0.010 & & 1.028(4)   \\ \hline
$32$ & 2.28 & 0.004 & 0.03 & 0.760(2)\\
& & 0.006 & &  0.765(2)\\
& & 0.008 & &  0.766(3)
\end{tabular}
\end{center}
\caption{\small\it Lattice data for the $\Omega$ baryon taken from ref.~\cite{Aoki:2010dy}.
\label{tab:data3}
}
\end{table}

A last comment is in order concerning the determination of the topological susceptibility in ref.~\cite{Chiu:2008jq}, based on gauge configurations for a smaller volume $(16)^3\times 32$ in refs.~\cite{Allton:2007hx,Aoki:2007xm}. In principle, this work could provide valuable additional information, but we have not been able to obtain consistent fits of the masses and decay constants of pions and kaons
with the three ensembles $(16)^3\times 32$, $(24)^3\times 64$, $(32)^3\times 64$, leading us to suspect an underestimation of the errors attached to the data for $(16)^3\times 32$. We have thus decided to keep only data for $(24)^3\times 64$, $(32)^3\times 64$, which were obtained for larger volumes and lighter quark masses, and thus less likely to be affected by sizable systematics.

\section{Finite-volume effects} \label{app:XiP}

As discussed in refs.~\cite{Luscher:1985dn,Leutwyler:1992yt,Becirevic:2003wk,Colangelo:2003hf,Colangelo:2004xr} and recalled in Sec.~\ref{sec:finvol}, 
for simulations where the time direction is much larger than the spatial ones,
the finite-volume effects 
amount at NLO to a modification of the chiral (tadpole) logarithms. One can make contact with the infinite-volume integral (and the corresponding tadpole logarithm)
in the following way, see eq.~(\ref{eq:XiP1}):
\begin{equation}
\Xi_P=\frac{\sigma_P}{L^3}-\frac{\mo_P^2}{8\pi^2}\log\frac{\mo_P^2}{\mu^2}=\xi_{1/2}(L,\mo_P^2)\,.
\end{equation}

We have already analysed this expression in ref.~\cite{DescotesGenon:2004iu}. $\xi_{1/2}$ contains a divergence of the form $1/(L^3\sqrt{\mo_P^2})$ in the chiral limit. However, one expects $\sigma_P$ to diverge like $1/(L^3 M_P)$ due to the propagation of pion zero modes, and in ref.~\cite{DescotesGenon:2004iu}, the alternative definition:
\begin{equation}\label{eq:XiP2}
\Xi_P=\left[\xi_{1/2}(L,\mo_P^2)-\frac{1}{L^3\sqrt{\mo_P^2}}\right]+\frac{1}{L^3M_P}\,,
\end{equation}
was proposed to reduce the size of HO corrections to the chiral expansions (the term in square brackets is analytic in quark masses) and it was used to determine the contribution of finite-volume effects in Resummed $\chi$PT for pseudoscalar masses and decay constants. 

In the present article, we define the HO remainders of the chiral expansions in the limit $L\to\infty$, and we assumed that even at finite volume, the variation of these remainders with respect to the volume would be very small, considering that the simulations are performed in the $p$-regime where $2mB_0L^2\gg 1$~\cite{Gasser:1987zq,Gasser:1987ah,Gasser:1986vb}. This is the reason for not including a volume-dependence on our expressions for the HO remainders. In a similar spirit, the choice between eqs.~(\ref{eq:XiP1}) and (\ref{eq:XiP2}) amounts to a redefinition of what would be considered as HO terms, for instance:
\begin{equation}
\tilde{e}_\pi \to \tilde{e}_\pi-\frac{1}{4\tilde{F}_\pi^2L^3}\left\{4\left[\frac{1}{M_\pi}-\frac{1}{\sqrt{\mo_\pi^2}}\right]+2\left[\frac{1}{M_K}-\frac{1}{\sqrt{\mo_K^2}}\right]\right\}\,,
\end{equation}
inducing only a small numerical correction of the acceptable range of variation for these remainders in the $p$-regime. This would not be the case in the $\epsilon$-regime, where  $2mB_0L^2\ll 1$,  which deserves a separate study~\cite{wip} and where the distinction between eqs.~(\ref{eq:XiP1}) and (\ref{eq:XiP2}) could induce large differences in the acceptable range for HO remainders.

Indeed, we have checked on the fits considered in the present paper that the outcome of finite-volume corrections according to eqs.~(\ref{eq:XiP1}) and (\ref{eq:XiP2})
led to very similar results if we allow reasonably large ranges of variation for the HO remainders. For convenience, we quote only the results obtained with eq.~(\ref{eq:XiP1}), using for HO remainders the dimensional estimates described in ref.~\cite{Bernard:2010ex}.

\end{document}